\documentclass[fleqn,usenatbib]{mnras}

\usepackage{newtxtext,newtxmath}

\usepackage[T1]{fontenc}

\DeclareRobustCommand{\VAN}[3]{#2}
\let\VANthebibliography\thebibliography
\def\thebibliography{\DeclareRobustCommand{\VAN}[3]{##3}\VANthebibliography}

\usepackage{graphicx}	
\usepackage{amsmath}	
\usepackage[normalem]{ulem}

\title[Cool outflows in MaNGA]{Cool outflows in MaNGA: a systematic study and comparison to the warm phase}

\author[C. R. Avery et al.]{Charlotte R. Avery$^{1}$\thanks{E-mail: c.r.avery@bath.ac.uk},
Stijn Wuyts$^{1}$,
Natascha M. Förster Schreiber$^{2}$, 
Carolin Villforth$^{1}$, \newauthor 
Caroline Bertemes$^{3}$, 
Stephen L. Hamer$^{1}$, 
Raman Sharma$^{1}$,
Jun Toshikawa$^{1}$, 
and Junkai Zhang$^{1}$ 
\\
$^{1}$Department of Physics, University of Bath, Claverton Down, Bath, BA2 7AY, UK\\
$^{2}$Max-Planck-Institut für Extraterrestrische Physik, Giessenbachstr. 1, D-85748 Garching, Germany\\
$^{3}$Zentrum für Astronomie der Universität Heidelberg Astronomisches Rechen-Institut, Mönchhofstr 12-14, D-69120 Heidelberg, Germany
}

\date{Accepted XXX. Received YYY; in original form ZZZ}

\pubyear{2015}

\begin{document}
\label{firstpage}
\pagerange{\pageref{firstpage}--\pageref{lastpage}}
\maketitle

\begin{abstract}
This paper investigates the neutral gas phase of galactic winds via the Na I D$\lambda \lambda 5890,5895$\AA \ feature within z $\sim$ 0.04 MaNGA galaxies, and directly compares their incidence and strength to the ionized winds detected within the same parent sample. We find evidence for neutral outflows in 127 galaxies ($\sim 5$ per cent of the analysed line-emitting sample). ${\rm Na\ I\ D}$ winds are preferentially seen in galaxies with dustier central regions and both wind phases are more often found in systems with elevated SFR surface densities, especially when there has been a recent upturn in the star formation activity according to the $\rm SFR_{5 Myr}/SFR_{800 Myr}$ parameter. We find the ionized outflow kinematics to be in line with what we measure in the neutral phase. This demonstrates that, despite their small contributions to the total outflow mass budget, there is value to collecting empirical measurements of the ionized wind phase to provide information on bulk motion in the outflow. Depending on dust corrections applied to the ionized gas diagnostics, the neutral phase has $\sim1.2-1.8$ dex higher mass outflow rates ($\dot{M}_{\rm out}$), on average, compared to the ionized phase. We quantify scaling relations between $\dot{M}_{\rm out}$ and the strengths of the physical wind drivers (SFR, $L_{\rm AGN}$). Using a radial-azimuthal stacking method, and by considering inclination dependencies, we find results consistent with biconical outflows orthogonal to the disk plane. Our work complements other multi-phase outflow studies in the literature which consider smaller samples, more extreme objects, or proceed via stacking of larger samples.

\end{abstract}




\section{Introduction}

Galaxy feedback is an integral part of galactic evolution and is a necessary ingredient in simulations to reduce the efficiency of star formation within galaxies and to recreate the observed shape of the galaxy stellar mass function \citep[e.g.,][]{Benson2003, Dutton2009, Dave2011, Somerville2015, Naab2017}. Dominant sources of feedback include the energy injection to the interstellar medium (ISM) from supernovae explosions \citep{Dekel1986, Efstathiou2000, Keller2015}, stellar winds \citep{Murray2005, Hopkins2014, Agertz2015}, and centrally accreting supermassive black holes known as Active Galactic Nuclei (AGN; see \citealp{Fabian2012} for a review on AGN feedback). It is widely accepted that AGN are particularly important for regulating galaxy growth at high masses. One mechanism by which AGN `feed back' into their surrounding environment is through the release of powerful jets which can extend well into the circumgalactic medium heating the halo gas, thus preventing it from cooling onto the disk. Another observed feedback phenomenon comes in the form of outflows which originate from AGN accretion and/or actively star-forming regions. These `galactic winds' are an example of an empirical signature of feedback in action and they can impact the ISM gas by displacing it and altering its thermal state, potentially impacting its ability to continue forming stars. Galactic winds are thus thought to be critical in shaping the evolution of galaxies across cosmic time \citep[see, e.g., review by][]{Veilleux2005}, and many studies have focused their efforts towards detecting and quantifying the characteristics of galactic winds from observations in an attempt to uncover their impact on host galaxy growth. 

Traditionally the focus of such studies has been on extreme objects, such as ultraluminous infrared galaxies (ULIRGs) and luminous AGNs, because the signatures of systematic blueshifts or broad velocity components tend to be most pronounced in them \citep[e.g.,][]{Heckman2000, Rupke2002, Schwartz2004, Martin2005, Rupke2005, Rupke2017}. Where efforts have been made to access the more normal low-redshift galaxy population, this has fruitfully been pursued using stacking of single-fibre SDSS spectra \citep{Chen2010, Cicone2016, RobertsBorsani2019, Concas2019}. However, by construction such measurements are restricted to the light captured within a single (in the case of SDSS, 3" diameter) fibre which samples different fractions of the galaxy depending on galaxy distance and size. Moreover, the galaxy parameter space that is potentially of influence to the presence and strength of outflow signatures is multi-dimensional and stacking approaches therefore inevitably average over a distribution of galaxy properties, even if sample statistics are sufficiently large to bin the population in more than one dimension. Among physical conditions potentially of relevance are intrinsic properties related to the activity and mode of star formation, and the presence and strength of any nuclear AGN activity. Dust shielding of the tracer element\footnote{The existence of neutral or molecular particles can be challenged when exposed to high-energy, ionizing photons. Dust grains can act to absorb these photons thus preventing photoionization.}or extinction effects may further render observable wind signatures more or less easily detectable. In addition, observational considerations such as our viewing angle with respect to the galaxy system or the portion of the galaxy covered by the spectroscopic observations may affect outflow detection and characterisation. 

In \citet{Avery2021} we presented a comprehensive study of the incidence and scaling relations of ionized gas outflows in a large sample of typical nearby galaxies probed by the MaNGA integral-field spectroscopic galaxy survey \citep{Bundy2015}. We made use of the spatially resolved nature of the IFU observations to (i) identify weak AGNs, (ii) remove the velocity field of systemic gas to better disentangle narrow (disk) and broad (outflow) components to the ionized line emission, (iii) estimate the extent of the outflows, and from this take outflow measurements within apertures best sampling the extent of the outflow for each individual object. The main caveat in our work being that we only probed the ionized gas phase of the outflows. 

The ionized phase, although being the easiest to constrain empirically, only makes up a small fraction of the full mass budget of the winds being driven from galaxies. Detailed multi-phase analyses of galactic outflows have been achieved in individual well-studied objects at low redshift \citep[e.g.][]{Feruglio2015, Leroy2015, Perna2019, Perna2020}. Studies of the multi-phase nature of outflows using larger samples tend to focus on systems which are at the extreme end in terms of their star formation rates \citep[SFRs;][]{Fluetsch2021}. Statistical studies for the more typical galaxy population rely on stacking techniques \citep{Sugahara2017, Concas2019, RobertsBorsani2020}. We therefore decide to extend the literature by presenting this counterpart to the ionized gas study by \citet{Avery2021}. In this paper we analyse the incidence of neutral gas outflows via the Na I D$\lambda \lambda 5890, 5895 \ \text{\AA}$ transition, starting from the same sample of line-emitting MaNGA galaxies that represent the underlying population considered in \citet{Avery2021}. 

After outlining the methodology of outflow identification and characterization in Section\ \ref{method.sec}, we discuss in Section\ \ref{results.sec} the relative incidence of detectable neutral gas winds compared to the subset of 322 objects showing evidence for ionized gas outflows, and the underlying analysed sample. For the 74 objects which have detectable winds in both gas phases, we contrast the ionized and neutral wind properties. For the full sample of 127 galaxies featuring detectable neutral gas outflows we quantify their strength as a function of key drivers such as the galaxies' star formation rate. We further  consider constraints on the outflow geometry by mapping the average spatial distribution of outflow via a radial/azimuthal stacking procedure, and by considering trends with inclination. We summarize our findings in Section\ \ref{summary.sec}.

Throughout the paper, we adopt a \citet{Chabrier2003} IMF and a flat $\Lambda$CDM cosmology with $\Omega_{\Lambda} = 0.7$, $\Omega_m = 0.3$ and $H_0 =70\ \rm{km}\ \rm{s}^{-1}\ \rm{Mpc}^{-1}$.

\section{Method and Sample Selection}
\label{method.sec}

\subsection{Extraction of galaxy spectra}
\label{sec:spectra}

This work makes use of data cubes from the MaNGA integral-field spectroscopic galaxy survey \citep{Bundy2015}. We refer the reader to Section 2.1 in \citet{Avery2021} for details on the parent sample used for this work. To summarize, we take the 4239 MaNGA Data Release 15 (DR15) objects which are successfully analysed by the MaNGA data analysis pipeline and cross-matched to the MPA-JHU database \citep{Kauffmann2003, Brinchmann2004, Salim2007}, providing measurements of total stellar masses and SFRs as well as their associated uncertainties. We created high signal-to-noise (S/N) spectra by stacking spaxels within elliptical apertures with major axis radii equal to 0.5$R_{\rm e}$, $R_{\rm e}$, and 1.5$R_{\rm e}$ for each individual object.\footnote{For reference, the typical spatial resolution is ${\rm FWHM} \sim 0.35 R_{\rm e}$ and $\sim 0.55 R_{\rm e}$ for sources in the Primary and Secondary MaNGA samples, respectively \citep[see][]{Wake2017}.} Here, $R_e$ is the galaxy's elliptical Petrosian effective radius in the $r$-band as measured from the NASA-Sloan Atlas (NSA) imaging and provided as output by the MaNGA data analysis pipeline. Before spaxel spectra were combined, the gaseous velocity field was removed and spaxel spectra were interpolated over a common velocity grid \citep[see also Section 2.2.1 in][]{Avery2021}. Errors on the combined spectra were calculated following equation (9) of \citet{Law2016} in order to account for covariance between spatially adjacent pixels.

\subsection{The analysed sample}
\label{sec:analysed_sample}

In \citet{Avery2021}, the ionized gas emission lines were analysed and modelled with a narrow Gaussian component tracing the systemic gas associated with the galaxy disk, and, where present, an additional broad Gaussian component tracing the outflowing gas.  This was done for each spectrum obtained within an elliptical aperture, if the S/N was found to be greater than $10$ in all lines which are used to perform the [NII]-BPT diagnostic (H$\beta$, [OIII]$\lambda 5007$, [N II]$\lambda$6583, H$\alpha$; see \citealp{Baldwin1981}).  A total of 2744 galaxies were found to have spectra obtained from at least one of their elliptical apertures satisfying this S/N criterion. These galaxies were dubbed the `analysed sample'. When referring in this paper to the `spectra within the analysed sample', we thus mean all spectra from the 0.5$R_{\rm e}$/$R_{\rm e}$/1.5$R_{\rm e}$ apertures which satisfy the S/N criterion. Of the galaxies making up the analysed sample, 322 showed evidence for hosting an ionized outflow from the presence of a broad-velocity component to the line emission. We refer the reader to Section 2.2 in \citet{Avery2021} for an in depth description of the methodology used to identify ionized outflows in the MaNGA sample. In this paper, we take the same high S/N spectra which we used to look for ionized outflows, to also search for outflows in the neutral phase using the ${\rm Na\ I\ D}$ spectral feature. For reference, galaxies in the analysed sample span a mass range $8 \lesssim \log(M_{\rm \star}) \lesssim 12$ (median $\langle \log(M_{\rm \star}) \rangle = 10.1$) and redshift range $0 \lesssim z \lesssim 0.15$ (median $\langle z \rangle = 0.03$).

We extended our analysed sample to include spectra which were obtained by stacking spaxels within a radius of 0.25$R_{\rm e}$ (corresponding roughly to the central resolution element). Again, we only considered spectra with S/N $> 10$ in all [NII]-BPT lines. We choose to only include the spectra with prominent emission line features because, as explained in Section \ref{sec:NaID_fitting}, we make use of the ionized gas kinematics extracted from a simultaneous fit to the H$\beta$, [OIII], H$\alpha$, [NII] and [SII] emission lines (as described in \citealp{Avery2021}) to place kinematic constraints on the doublet component arising from systemic ISM gas in the Na I D feature.

\subsection{Na I D profile fitting}
\label{sec:NaID_fitting}

We search for gaseous outflows in the neutral phase by modelling the Na I D profile with a combination of its constituent components. Where an outflow is present in front of the disk along our line of sight, we expect to see a corresponding blue-shifted doublet in absorption in the Na I D profile (see Fig. 1 in \citealp{RobertsBorsani2019}). This is due to the outflow being illuminated by the background galaxy disk.

Although outflowing gas can plausibly re-emit photons at Na I D wavelengths, due to the isotropic re-emission of photons, any emission that reaches us will be overcome by the absorption. Furthermore, we expect the re-emitted light from the receding outflow to be substantially dimmed by intervening dust in the galaxy disk. For these reasons we expect any re-emission in the outflow to be difficult to detect, and by considering the noise level in our data we choose not to account for this effect in our models. We therefore search for neutral outflows solely based on a blue-shifted absorption signature in ${\rm Na\ I\ D}$.

Unlike for the ionized emission, where we can use individual unblended lines (e.g., H$\beta$, [OIII]$\lambda 4959$, [OIII]$\lambda 5007$) to aid the profile decomposition into disk (ISM) and outflow components (see Section 2.2 in \citealp{Avery2021}), the Na I D feature is a heavily blended line complex: the observed Na I D feature has contributions from an absorption doublet arising in stellar atmospheres, doublet emission or absorption stemming from gas clouds at systemic velocities in the ISM, and (possibly) a contribution from wind gas moving at larger velocities. A careful spectral decomposition into stellar and gaseous components is thus critical for a comprehensive modelling of the ${\rm Na\ I\ D}$ line complex. 

To this end, we make use in this work of the Penalized Pixel-Fitting (\texttt{ppxf}) code by \citet{Cappellari2017} to find a best fitting stellar model for each spectrum in the analysed sample. We perform the stellar fitting on the full wavelength range provided by MaNGA whilst masking the Na I D feature since, as noted above, we expect a significant contribution to the Na I D feature from gas in a number of cases, and we do not want this gas to wrongfully influence the stellar models used. We refer the reader to Section of 2.2.1 of \citet{Avery2021} for further details on the \texttt{ppxf} settings used.

For each of the high S/N spectra in the analysed sample, and their associated stellar models, we extracted the wavelength region containing the Na I D $\lambda \lambda 5890, 5895 \textrm{\AA}$ doublet feature. Whilst fitting the Na I D feature, we work in velocity space within a velocity window of $\pm 2,000 \rm \ km \ s^{-1}$ centered on the Na I D$\lambda 5890$ line. With the stellar model in hand, we use \texttt{mpfit} (\citealp{mpfit}, updated for Python by Sergey Koposov) to perform a decomposition on the Na I D gas component. Given that the He I $\lambda 5876$ emission feature is often blended with the Na I D feature, we decided to model this feature simultaneously with Na I D using a single Gaussian. We fix the kinematics of the He I line emission to that derived for the strong-line ionized gas emission (H$\beta$, [OIII], H$\alpha$, [NII], [SII]) and leave the line amplitude ($A_{\rm He I}$) free to fit.

\begin{figure*}
\centering
\includegraphics[width=0.49\textwidth]{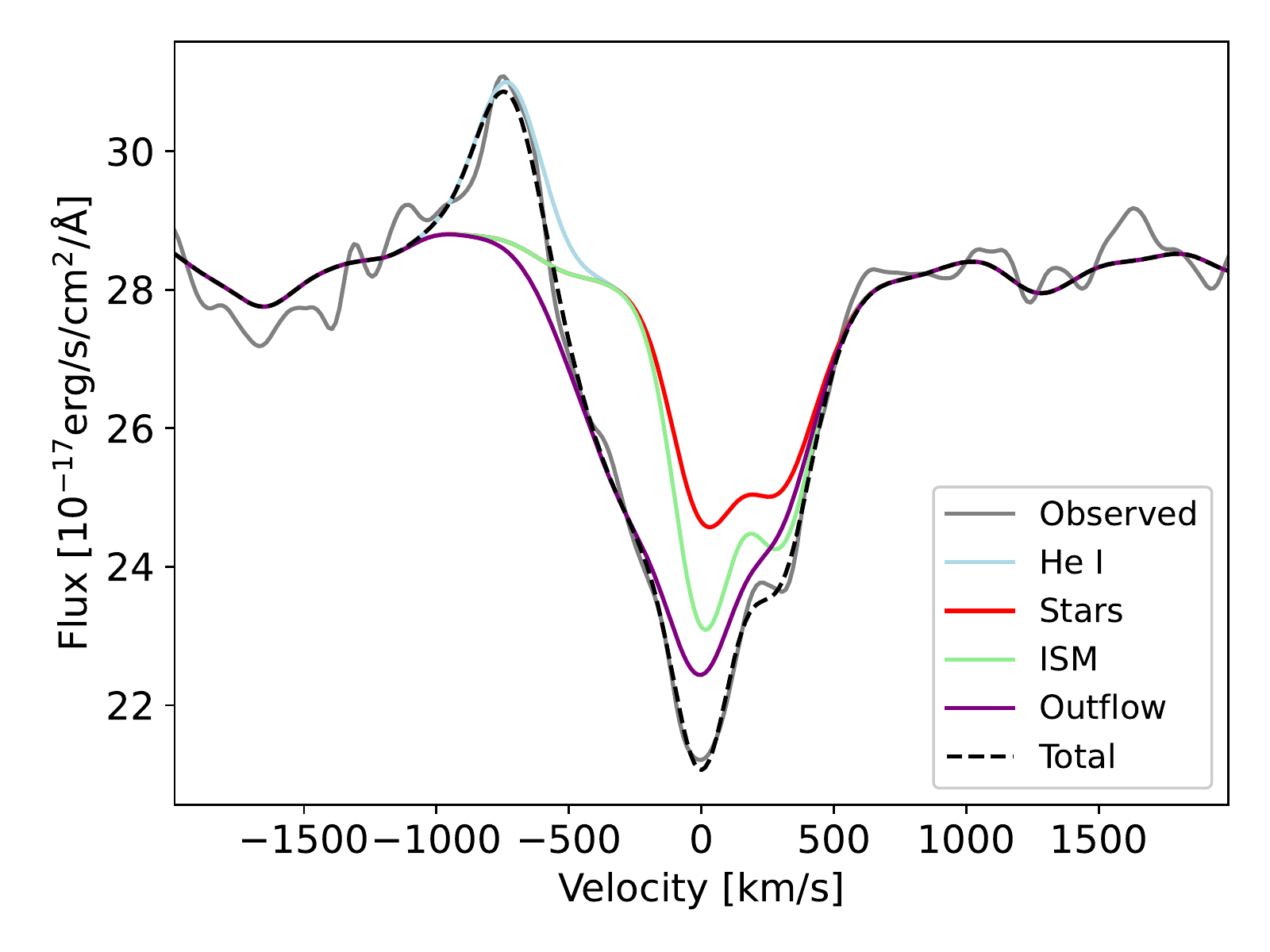}
\includegraphics[width=0.49\textwidth]{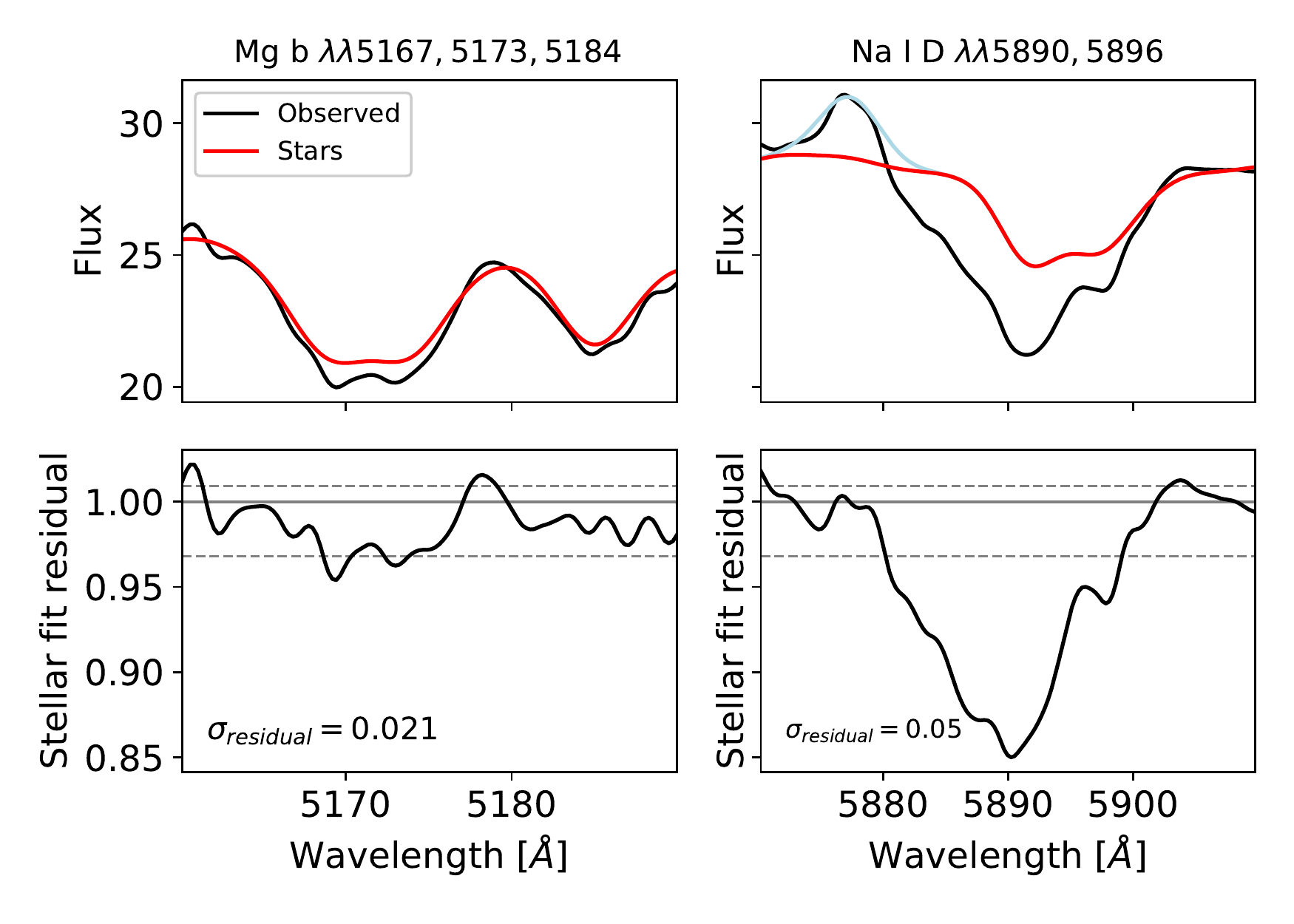}
\caption{{\it Left:} Example Na I D multi-component profile fitting for a MaNGA object. The total best fitting profile is shown with the black dashed line. {\it Right:} Upper panels show the best fitting \texttt{ppxf} stellar spectrum in the wavelength regions containing the Mg $b$ triplet and the Na I D doublet. Lower panels show the residuals from the removal of the stellar model (and He I model). If a gas component is present, we expect significantly larger residuals in Na I D compared to Mg $b$, as shown here.}
\label{fig:profile_fit}
\end{figure*}

To model the Na I D gas component, we follow the fitting procedure outlined by \cite{Baron2021}, which is based on the completely overlapping atoms model scenario formulated by \cite{Rupke2005}. To determine if a neutral gas outflow is present, each spectrum in the analysed sample is fit with a simple ISM-only model and an ISM + outflow model. To account for the fact that the ISM component of the Na I D feature can either be emitting or absorbing (e.g., see \citealp{Concas2019}), we consider four types of models in total: (i) ISM absorption + outflow, (ii) ISM emission + outflow, (iii) ISM absorption, (iv) ISM emission. We then compare the different models using the Bayesian information criterion (BIC) statistic. The model was performed on the observed spectrum in velocity space $F(v)$. Below we describe the details of each model.

Since emission terms are additive and absorption terms are multiplicative, we use the following equation for model (i):
\bigskip

(i) $F(v) = [F_{\textrm{HeI}}(v) + F_{\textrm{stars}}(v) + c] \times F_{\textrm{NaID, abs}}(v)$
\bigskip

Here, $F_{\textrm{HeI}}(v)$ represents the He I Gaussian model, $F_{\textrm{stars}}(v)$ the stellar model (determined using \texttt{ppxf}), and $c$ represents a free constant which can be considered a nuisance parameter to account for ‘bad’ continuum fitting in the Na I D region. Note that we find typical values of $c$ to be small, at the level of 0.5 per cent of the stellar continuum. $F_{\textrm{NaID, abs}}$ is the Na I D gas component model, and is given by the following equation:
\begin{equation}
\begin{aligned}
    F_{\textrm{NaID, abs}}(v) = 1 \ - \ C_f \ + \ C_f \textrm{exp} \{- \tau_{\rm ISM,b}(v) - \tau_{\rm ISM,r}(v) \\
    - \tau_{\rm B,b}(v) - \tau_{\rm B,r}(v)\}
\label{eqn:NaID_abs}
\end{aligned}
\end{equation} 

$C_f$ is the covering fraction of the gas. $\tau_{\rm b}(v)$ and $\tau_{\rm r}(v)$ are the optical depths of the blue (NaID $\lambda 5890$) and red (NaID $\lambda 5895$) doublet components, where the subscript `ISM' refers to optical depths associated with the absorbing ISM gas, and `B' refers to the absorbing gas making up the broad (outflow) component. The optical depth is given by the following equation:
\begin{equation}
    \tau(v) = \tau_0\ \textrm{exp} (- \Delta v^2 / b^2)
\end{equation}
where $\tau_0$ is the optical depth at the center of the line, $\Delta v$ is the velocity shift of the line center from the galaxy's systemic velocity, and $b$ is the Doppler parameter. $b$ is related to the velocity dispersion of the line by $b = \sqrt 2 \sigma$. 

Both $\Delta v$ and $b$ are tied between the red and blue line components of the doublet features arising from the ISM and outflow absorption. To prevent the ISM model parameters  taking unrealistic values, we place the following constraints on the velocity dispersion and velocity shift of the Na I D ISM absorption component: $\sigma_{\rm ISM} \leq \sigma_{\rm N}$ and $\Delta v_{\rm ISM} \leq 2\sigma_{\rm N}$, where $\sigma_{\rm N}$ represents the width of the narrow component in the ionized line emission.

The optical depth at line center for the red doublet component is left to fit freely for both the ISM ($\tau_{\rm 0,r,ISM}$) and outflow ($\tau_{\rm 0,r,B}$) components in the model. We determine the blue line center optical depths ($\tau_{\rm 0,b,ISM}$ and $\tau_{\rm 0,b,B}$) using the doublet ratio method \citep{Somerville1988}.

In total, model (i) has 9 free parameters: $A_{\rm He I}$,  $C_f$, $\tau_{\rm 0,r,ISM}$, $\Delta v_{\rm ISM}$, $b_{\rm ISM}$,  $\tau_{\rm 0,r,B}$, $\Delta v_{\rm B}$, $b_{\rm B}$, $c$. 

To account for galaxies with an emitting ISM gas component we use the following model:
\bigskip

(ii) $F(v) = [F_{\textrm{He I}}(v) + F_{\textrm{stars}}(v) + c + F_{\textrm{NaID, em}}(v)] \times F_{\textrm{NaID, abs}}(v)$ 
\bigskip

where the ISM component $F_{\textrm{NaID, em}}(v)$ is described by a Gaussian doublet with kinematics fixed by the narrow component model to the ionized line emission. The amplitude of the red doublet line $A_{\rm r, ISM}$ is left as a free parameter and the ratio of the doublet lines $A_{\rm r, ISM}/A_{\rm b, ISM}$ is limited within the range [1,2].

The outflow component is defined as follows:
\begin{equation}
\begin{aligned}
    F_{\textrm{NaID, abs}}(v) = 1 - C_f + C_f \textrm{exp} \{- \tau_{\rm B,b}(v) - \tau_{\rm B,r}(v)\}
\end{aligned}
\end{equation}

In total, model (ii) has 8 free parameters: $A_{\rm He I}$, $A_{\rm r, ISM}$, $A_{\rm r, ISM}/A_{\rm b, ISM}$, $C_f$, $\tau_{\rm 0,r,B}$, $\Delta v_{\rm B}$, $b_{\rm B}$, and $c$.

(iii) Same as model (i) except the optical depths associated with the outflow doublet component ($\tau_{\rm B}$) are removed from the $F_{\rm NaID, abs}(v)$ factor, leaving 6 free parameters: $A_{\rm He I}$,  $C_f$, $\tau_{\rm 0,r,ISM}$, $\Delta v_{\rm ISM}$, $b_{\rm ISM}$, $c$. 

(iv) Same as model (ii) except the $F_{\rm NaID, abs}(v)$ factor is removed from the model, leaving 4 free parameters: $A_{\rm He I}$, $A_{\rm r, ISM}$, $A_{\rm r, ISM}/A_{\rm b, ISM}$, $c$.

Errors on fit parameters were determined from the covariance matrix returned by \texttt{mpfit}, which took into account the observed error spectrum (see Section \ref{sec:spectra}) in the fitting process.

\subsection{Outflow criteria}
\label{sec:criteria}

For a galaxy to show evidence for neutral gas outflows, all of the following criteria need to be satisfied in at least one of the aperture sizes considered:

\begin{enumerate}
    \item[(a)] Either model (i) or (ii), in other words a model with an outflow component, must provide a significantly better fit to the observed spectrum over both ISM-only models [(iii) and (iv)] according to the BIC criterion. We choose the BIC criterion as it prevents selection of an over-fit model by penalising models with a higher number of free parameters. For two models, 1 and 2 with total chi-squared $\chi_{\rm 1,2}$, $p_{\rm 1,2}$ free parameters and $n$ data points, $\Delta \textrm{BIC} = \chi^2_{\rm 2} + p_{\rm 2} \textrm{ln}(n_{\rm 2}) - \chi^2_{\rm 1} - p_{\rm 1} \textrm{ln}(n_{\rm 1})$. We take $\mathrm{\Delta BIC > 10}$ to be evidence that an ISM+outflow model is required over either ISM-only models.

    \item[(b)] The absolute value of the peak amplitude of the outflow component must be at least 3 times larger than the standard deviation of the flux within continuum windows surrounding the Na I D line.

    \item[(c)] The blueshift of the outflow component from the ISM component must be less than -50 $\rm km \ s^{-1}$ (i.e., in amplitude larger than 50 $\rm km\ s^{-1}$) to ensure the components are kinematically distinct (see also \citealp{Rupke2005b}), but also greater than $- 640 \ \rm km \ s^{-1}$ to prevent components from drifting off and fitting noise.

    \item[(d)] The Doppler parameter of the outflow component ($b_{\rm B}$) must be larger than the Doppler parameter of the ISM gas.\footnote{Given the definition $b = \sqrt 2 \sigma$, this is equivalent to the requirement that the velocity dispersion of the outflow component must exceed that of the ISM gas.} Furthermore, to avoid the outflow component fitting for any continuum residuals that may not have been accounted for properly, we impose the upper limit: $b_{\rm B} < 2000 \ \rm km \ s^{-1}$.

    \item[(e)]  Na I D must contain a significant gas component. To check for this, we compare the standard deviation of the flux residuals remaining after the removal of the stellar model in the Na I D and Mg I $\lambda \lambda 5167, 5173, 5184$ (Mg $b$) wavelength regions, where, unlike Na I D, the Mg $b$ triplet only has absorption associated with stellar atmospheres.\footnote{Due to the similarity between the production of ${\rm Mg}\ b$ and ${\rm Na\ I\ D}$ stellar transitions, and their similar ionizing potentials, the Mg $b$ absorption is a good probe for the amount of stellar absorption expected in Na I D. Given that the robust extraction of the Na I D gas component relies on a good stellar fit, a comparison of the stellar residuals in the two regions is an adequate way to ensure our Na I D gas model is securely tracing features arising from the ISM/outflowing gas, and not the residuals of a poor continuum fitting.} We require the standard deviation of the residual in Na I D to be at least $10$ per cent larger than the residual in ${\rm Mg}\ b$. 
    
    \item[(f)] The spectral fitting must pass a visual inspection, where we fail objects if the presence of an excess flux residual in Na I D is ambiguous, after the removal of the stellar component.
\end{enumerate}

Applying the criteria in the order that they are listed above limited the number of potential wind galaxies down from 2744 galaxies to 1252, 579, 217, 173, 147, and finally to 127. 

The lower limit value in criteria (c) was chosen by adjusting the value until we were satisfied, by eye, that there were no false positives due to the fitting of noise in the outflow sample. The upper limit in criteria (d) was also chosen based on the same reasoning. Criterion (e) was chosen based on an eyeball inspection of the residuals from the stellar fitting in the Mg $b$ and Na I D wavelength regions (i.e., by inspecting plots equivalent to left panel of Fig. \ref{fig:profile_fit}). We highlight that these somewhat arbitrary choices of criteria only eliminate a handful of objects from the outflow sample, and when applying criteria (c) and (d), it is the upper and lower limits respectively which have the majority impact. We consider the impact of the $-50 \ \rm km s^{-1}$ upper limit imposed in criterion (c). This is modestly below the typical MaNGA spectral resolution of $\sim 70  \ \rm km s^{-1}$, which we believe is appropriate as accounting for the line spread function allows for centroid determination at this level, and given that we are able to constrain upper limits on the kinematics of the Na I D-ISM component based on the emission line fitting. Furthermore, we find that $\sim 80$ per cent of detected Na I D winds have an outflow component which is blue-shifted by more than $70\ \rm km\ s^{-1}$.

In 37 cases, both models (i) and (ii) satisfy the above outflow criteria. For these systems, we then choose the better fitting model out of (i) and (ii) as the one with the largest associated $\Delta$BIC value. 

An example decomposition of the Na I D profile for a galaxy showing evidence of a galactic wind in the neutral gas phase is shown in the left-hand panel of Fig. \ref{fig:profile_fit}. For this object, we found model (i) to provide the best fit to the observed line profile. In the right-hand panel of Fig. \ref{fig:profile_fit}, where we show a comparison of the residuals after the removal of the stellar spectrum in the wavelength regions containing the Mg $b$ and the Na I D feature, we can see a significant absorption residual in the Na I D wavelength regime compared to the Mg $b$ region, supporting our gas-model determination.

\begin{table} 
{ 
\begin{tabular}{c c c} 
\hline 
Aperture size & Number of spectra & Number of outflows \\
 & analysed & detected \\
\hline 
$0.25 R_{\rm e}$   & 2192 & 76  \\ 
$0.5 R_{\rm e}$   & 2457 & 71 \\ 
$1 R_{\rm e}$   & 2658 & 64 \\ 
$1.5 R_{\rm e}$   & 2721 & 52 \\ 
\hline 
\end{tabular} 
} 
\caption{For each aperture size investigated, we list the number of spectra which were analysed in search of neutral gas outflows, and of these, the number of which showed evidence for neutral outflows. Detection rates are highest in smaller apertures. Most galaxies are analysed using more than one aperture size; there are 2744 unique galaxies which have analysed spectra, and 127 of these show evidence for outflows. }
\label{tab:aper}
\end{table} 

We find 127 galaxies in total that show evidence for neutral gas outflows within at least one of their 0.25/0.5/1/1.5 $R_{\rm e}$ stacks, based on the above criteria (see also Table\ \ref{tab:aper}). These galaxies are referred to as the `neutral outflow sample' in the remainder of this paper. Of these, we find 76 and 51 to be from non-active and active galaxies, respectively, based on the position of the line ratios determined within $0.25R_{\rm e}$ relative to the  \citet{Kauffmann2003} separation curve on the [NII]-BPT diagnostic diagram.

\subsection{Outflow extent and adopted aperture size}
\label{sec:Rout}

Similar to \citet{Avery2021}, for each object in the neutral outflow sample, we determine an optimal aperture size from which we extract outflow properties, where we take the optimal aperture as an ellipse with major axis radius equal to radial extent of the outflow ($R_{\rm out}$). For the ionized gas we performed an annular binning method to measure the radial extent of the broad velocity component (originating from the outflow) to the H$\alpha$ emission (see Section 2.2.6 in \citealp{Avery2021}). However, since the Na I D feature is relatively weak compared to the ionized emission, we find this method to be unsuitable due to the relatively low S/N in each of the radial bins. Instead we determine a rough estimate of the outflow extent as follows. For each object in the outflow sample, we consider which of the analysed apertures (out of 0.25$R_{\rm e}$, 0.5$R_{\rm e}$, 1$R_{\rm e}$, 1.5$R_{\rm e}$) shows evidence for outflow based on the criteria outlined in Section \ref{sec:criteria}. If outflows are seen in more than one aperture size for a given galaxy, we determine which of the apertures best samples the outflow light in that galaxy, and we take that aperture size as a measure of the radial extent of the outflow.\footnote{We do not simply choose the largest aperture size as including spaxels outside the outflow extent may simply have the effect of adding more continuum light, and thus more noise, to our spectra making the observed outflow absorption signature more difficult to detect.} We achieve this by adopting the outflow extent as the aperture size that maximises the product of the covering fraction of the gas, the column density of the gas, and the outflow velocity (see Section\ \ref{sec:outflow_params} for detailed parameter definitions). Henceforth, we refer to this aperture as the `outflow aperture', and all neutral outflow diagnostics for the object under consideration are quantified using the best-fit parameters from the decomposition of the Na I D feature of the spectrum extracted from this aperture.

Of the 127 objects which show evidence for neutral gas outflows, the number of outflows with an extent of either 0.25, 0.5, 1, or 1.5 $R_{\rm e}$ is listed in Table \ref{tab:Rout}. We find most outflows are centrally concentrated within $\sim 0.25 R_{\rm e}$. 

\begin{table} 
{ 
\begin{tabular}{c c c} 
\hline 
Outflow extent & Number of galaxies  \\
\hline 
$0.25 R_{\rm e}$ & 62  \\ 
$0.5 R_{\rm e}$ & 29 \\ 
$1 R_{\rm e}$ & 22 \\ 
$1.5 R_{\rm e}$ & 14 \\ 
\hline 
\end{tabular} 
} 
\caption{A table summarising the number of galaxies which have outflows reaching a given outflow extent. We estimate the outflow extent by considering which aperture size best encompasses the outflow gas based on the product of the covering fraction of the gas, the column density of the gas, and the outflow velocity.}
\label{tab:Rout}
\end{table} 

As a sanity check, we re-ran our analysis when the outflow radius within each galaxy is taken to be that aperture size which maximises the full equation\ \ref{eq:Mdot_out} for outflow rate (including the $R_{\rm out}$ factor itself, taken to be the radius of the aperture under consideration). We find that our key results do not change, and we recover the same $\dot{M}_{\rm out}$ dependencies on other galaxy properties as recorded in this paper.

\subsection{Outflow properties}
\label{sec:outflow_params}

This section describes how we quantify the physical characteristics of neutral gas outflows, where detected among the MaNGA galaxy population. In summary, we measure the properties of neutral gas outflows within individual galaxies using the parameters derived from the decomposition of the Na I D profile, obtained from the outflow aperture.

To be consistent with \citet{Avery2021}, we define the outflow velocity as follows:
\begin{equation}
    v_{\rm out} = | \Delta v_{\rm B} - 2\sigma_{\rm B} |
\label{eq:vout}
\end{equation}

where $\Delta v_{\rm B}$ is the shift of the line profile centroid of the outflow component relative to the galaxy's systemic velocity as defined from the narrow component emission in the ionized gas lines, and $\sigma_{\rm B}$ is the velocity dispersion of the outflow component. 

To estimate the mass outflow rate, we use the equations from \citet{Baron2021}. These are based on the model described by \citet{Rupke2005b} which makes several assumptions about the geometry of the wind and calculates the time-averaged outflow rates (as opposed to instantaneous outflow rates allowing for better comparisons with the SFR). We refer the reader to \citet{Rupke2005b} for model details.
\begin{equation}
    \dot{M}_{\rm out} = 11.45\ \textrm{M}_{\odot}\ {\rm yr}^{-1}\ \left(\dfrac{C_{\Omega}}{0.4}C_f\right) \left(\dfrac{N_{\rm H}}{10^{21}\rm \ cm^{-2}} \right) \left(\dfrac{R_{\rm out}}{1 \rm \ kpc}\right) \left(\dfrac{v_{\rm out}}{200 \ \rm km \ s^{-1}}\right)
\label{eq:Mdot_out}
\end{equation}

We assume a value of 0.4 for the large-scale covering factor $C_{\Omega}$, which is related to the wind’s opening angle. This follows the assumption made by \citet{Rupke2005b} for infrared galaxies. Covering fractions $C_f$ are computed for individual galaxies from our ${\rm Na\ I\ D}$ fits. We find a range of $C_f$ values spanning 0.05 to 1. The outflow radius $R_{\rm out}$ is estimated for each object as described in Section \ref{sec:Rout}. 
$N_{\rm H}$ is the column density of the gas and can be estimated using the equation: 
\begin{equation}
    N_{\rm H} = \dfrac{N_{\rm Na I}}{(1 - y)10^{A+B+C}}
\end{equation}

where we follow \citet{Baron2021} in defining the parameters as follows. The sodium neutral fraction $(1 - y) = 0.1$, the sodium abundance term $A = \textrm{log}(N_{\rm Na}$/N$_{\rm H}) = -5.69$, and the sodium depletion term $B = \textrm{log}(N_{\rm Na}$/N$_{\rm H, total}) -  \textrm{log}(N_{\rm Na}/N_{\rm H, gas}) = -0.95 $. We take $C$, the gas metallicity term, from the mass -- metallicity relation of \citet{Curti2020}. 

The column density of the sodium gas $N_{\rm NaI}$ can be estimated using our fitting results for the line center optical depth of the red doublet component \citep{Draine2011}:
\begin{equation}
    N_{\rm Na} = 10^{13} \mathrm{cm^{-2}} \left(\dfrac{\tau_{\rm 0,r}}{0.7580}\right) \left(\dfrac{0.4164}{f_{\rm lu}}\right)  \left(\dfrac{1215\text{\AA}}{\lambda_{\rm lu}}\right) \left(\dfrac{b}{10 \ \rm km \ s^{-1}}\right)
\end{equation}
where $f_{\rm lu} = 0.32$ and $\lambda_{\rm lu} = 5897$\AA \ are parameters describing the ${\rm Na\ I\ D}$ atomic transition. 

To obtain an estimate of the error on our $\dot{M}_{\rm out}$ measurement, we propagate errors on the dependent parameters $C_f$, $\tau_{\rm 0,r}$, $b$, $v_{\rm out}(\Delta v_{\rm B}, \sigma_{\rm B})$ and $R_{\rm out}$, where the uncertainty on $R_{\rm out}$ accounts for the finite spatial resolution, and uncertainties on the other parameters are taken from the output of the \texttt{mpfit} fitting process (Section \ref{sec:NaID_fitting}).

\section{Results and Discussion}
\label{results.sec}

In this section, using the outflow properties derived from the ${\rm Na\ I\ D}$ feature decomposition, we present our key findings and discuss their context within the relevant literature. First, we explore the incidence of galaxies with visible neutral winds among the underlying galaxy population and in relation to the ionized wind population (Section\ \ref{incidence.sec}). Second, for a subsample of wind galaxies, we are able to directly compare the kinematics and outflow energetics in different phases (Section\ \ref{phases.sec}). Third, we present scaling relations between the mass outflow rate and outflow drivers (Section\ \ref{scaling.sec}). Finally, we consider the geometry of the winds (Section\ \ref{geometry.sec}). 

\subsection{Outflow incidence}
\label{incidence.sec}

\subsubsection{Which line-emitting galaxies show Na I D outflow signatures?}

Of the 2744 line-emitting MaNGA objects analysed in our pipeline, 322 ($\sim12$ per cent) show evidence for warm ionized outflows appearing as broad velocity components to the ionized line emission, and 127 ($\sim5$ per cent) show evidence for cool neutral outflows based on the requirement of an additional blue-shifted absorption component to the Na I D feature. A subset of only 74 MaNGA galaxies show outflows in both the ionized and neutral gas phases. 

We find a significant fraction of outflows originate from galaxies with evidence for AGN activity based on their nuclear BPT classifications: $\sim 60$ per cent of ionized outflows and $\sim 43$ per cent of neutral outflows originate from active galaxies.

Among the parent analysed MaNGA sample, we find Na I D wind detections to be more prevalent among systems of higher stellar masses ($M_{\star}$), central stellar velocity dispersion (calculated within 1 $R_{\rm e}$, $\sigma_{\rm \star R_e}$), specific SFRs (sSFR), SFRs, star formation surface densities $\Sigma_{\rm SFR}$ ($\Sigma_{\rm SFR} \equiv \mathrm{SFR}/2\pi R_{\rm e}^{2}$), and AGN luminosities $L_{\rm AGN}$.\footnote{We refer the reader to Section 2.2.5 of \citet{Avery2021} for an in depth description of the calculation of AGN-luminosities used in this work. In short, we measure $L_{\rm AGN}$ by applying a bolometric correction to the AGN-photoionized [OIII] emission.} Furthermore, when compared to the ionized outflow sample, ${\rm Na\ I\ D}$ winds are preferentially seen towards higher sSFRs, SFRs and $\Sigma_{\rm SFR}$. We confirm this using the Kolmogorov-Smirnov (K-S) test, which quantifies the distance between cumulative distributions. We present histograms illustrating these effects in Appendix A. Among the parameters listed above, we find the $\Sigma_{\rm SFR}$ distribution to display the largest contrast when comparing the ionized and Na I D wind samples, with a K-S value of 0.24 (p-value $p = 3 \times 10^{-5}$).

As outlined in Section \ref{sec:analysed_sample}, we search for neutral winds among the same analysed sample of line-emitting galaxies as defined by \citet{Avery2021}, even though the detection of neutral winds is based on the ${\rm Na\ I\ D}$ {\it absorption} feature. Although the tell-tale, blue-shifted absorption signature of outflows could in principle be present in ${\rm Na\ I\ D}$ among the sample of galaxies which do not show significant line emission (and thus do not make it into the analysed sample), we expect this to be the case for only a very small number of objects. This is because the galaxies which do not satisfy our criteria for significant line emission have lower $\Sigma_{\rm SFR}$, when compared to the underlying MaNGA sample, whereas our results show that neutral outflows are prevalent among systems with enhanced $\Sigma_{\rm SFR}$ compared to the underlying population. We therefore assume the number of outflow detections in non-line-emitting galaxies to be negligible, and thus conclude incidence rates of $\sim 8$ per cent and $\sim 3$ per cent among the full MaNGA sample for the ionized and neutral winds, respectively.\footnote{Evaluated with respect to the full population of galaxies with $\log(M_*) \gtrsim 9$ in the nearby Universe, the fraction of objects with detectable wind signatures is anticipated to be yet lower, given that the MaNGA sample was drawn from a flat mass distribution rather than straight from the galaxy stellar mass function \citep{Wake2017}, and that outflow detections are more prevalent towards the high-mass end (see Appendix Figure\ \ref{fig:incidence}).}

\subsubsection{Impact of dust}

\begin{figure*}
\centering
\includegraphics[width=0.98\linewidth]{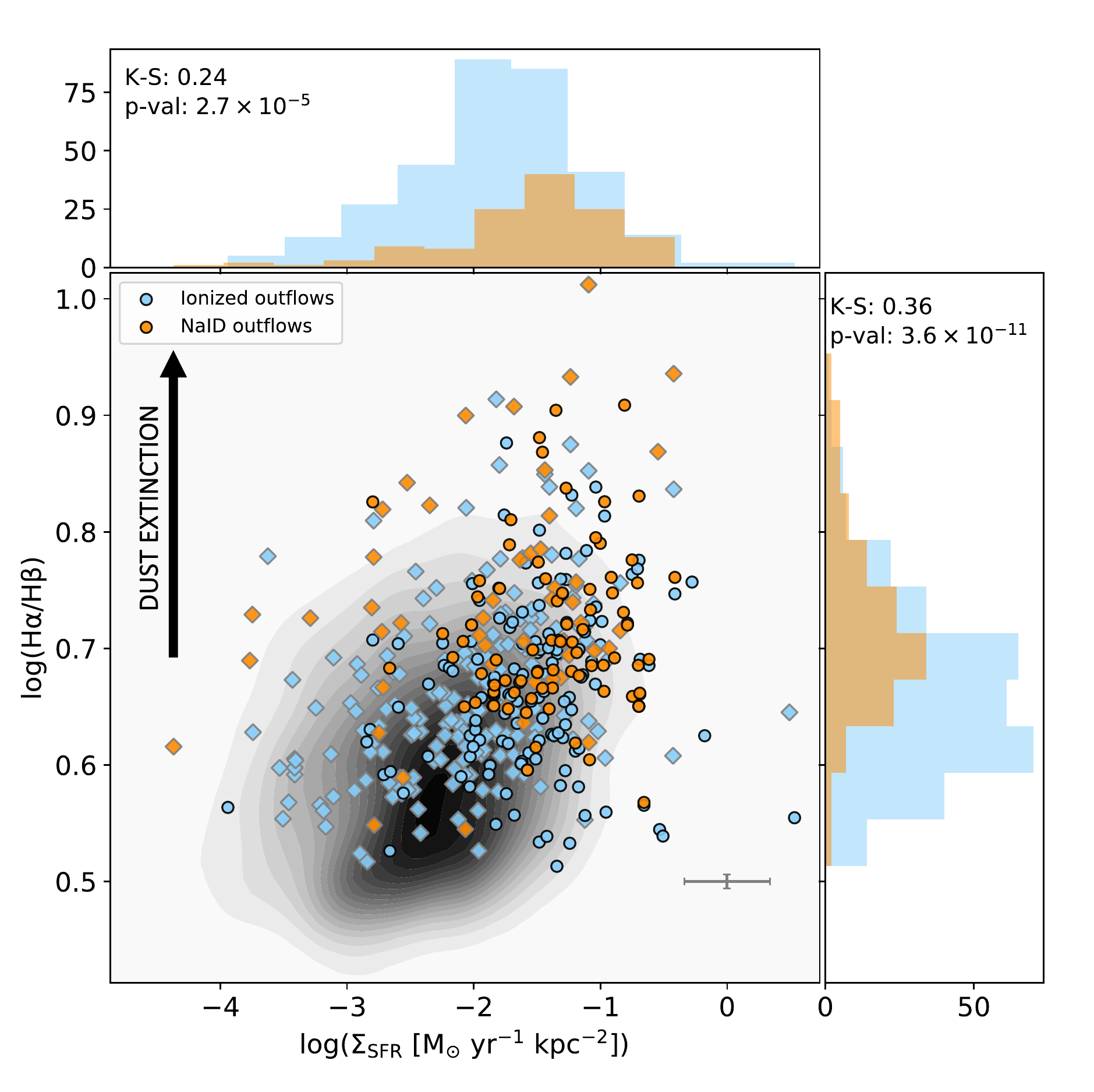}
\caption{Incidence of ionized (blue) and neutral (orange) gas outflows as a function of total galaxy star-formation surface density, and dust content probed by the Balmer decrement within the outflow aperture. The underlying analysed sample is shown in grey shades. Circle and diamond symbols denote galaxies without and with nuclear activity, respectively.  Neutral gas outflows are preferentially detected in more dusty systems. }
\label{fig:dust}
\end{figure*}

Although outflows detected with Na I D are offset towards higher $\Sigma_{\rm SFR}$ when compared to both the analysed sample and the ionized outflow sample, this offset can to a large degree be attributed to a higher incidence of neutral outflow detections in dustier systems. This effect is presented in Fig. \ref{fig:dust}. Here we show the incidence of ionized and neutral outflows in $\Sigma_{\rm SFR}$ and H$\alpha/$H$\beta$, where the former parameter quantifies the galaxy's global SFR surface density, and the latter parameter probes the amount of dust attenuation within the outflow aperture (of radius governed by the outflow extent; see Section \ref{sec:Rout}). It is evident that the Balmer decrement and $\Sigma_{\rm SFR}$ values are correlated.  Where star formation is more active, this tends to be associated with larger columns of dust.  \citet{Li2019} analyze this relation and document how the (substantial) scatter around it can be attributed to variations in metallicity, inclination, as well as dust geometry.  Of note to our investigation of outflow incidence, one can appreciate from Fig.\ \ref{fig:dust} that at fixed $\Sigma_{\rm SFR}$ the galaxies with ${\rm Na\ I\ D}$ winds sample preferentially the regime with $\log({\rm H}\alpha/{\rm H}\beta) \gtrsim 0.65$ (corresponding to $A_{V, {\rm gas}} \gtrsim 1.2$), whereas ionized winds are detected across a broader range of Balmer decrements, down to low obscuration levels ($\log({\rm H}\alpha/{\rm H}\beta) \sim 0.5$ corresponding to $A_{V,{\rm gas}} \sim 0.3$).  At such low obscurations, ionized outflows span the full range from high $\Sigma_{\rm SFR}$ (predominantly star formation driven winds) to low $\Sigma_{\rm SFR}$ (predominantly AGN driven winds). Note that star formation and AGN driven winds can be identified separately in Fig. \ref{fig:dust} as circle and diamond markers, respectively. At higher obscuration levels, no clear differentiation between $\Sigma_{\rm SFR}$ distributions of ionized versus neutral outflows is seen at fixed H$\alpha$/H$\beta$.  We thus conclude that it is primarily dust attenuation which dictates wind detectability in the neutral phase as probed by ${\rm Na\ I\ D}$.

This result is not unexpected given that dust grains could potentially provide the shielding from ionizing radiation that is necessary to sustain a significant abundance of detectable Na I D gas, given its relatively low ionizing potential compared to hydrogen.

Furthermore, a strong positive correlation between the equivalent width of the Na I D absorption and the Balmer decrement computed within the outflow aperture (Spearman's rank correlation coefficient $R_s = 0.6, p = 2\times10^{-13}$; see Fig. \ref{fig:EW_dust}) points towards dust playing an important role in Na I D detection. Previous studies have recognised this trend and have further shown that Na I D outflows themselves have a significant dust component \citep{Heckman2000, Rupke2005b, Veilleux2005, Chen2010, Veilleux2020}.

\begin{figure*}
\centering
\includegraphics[width=\linewidth]{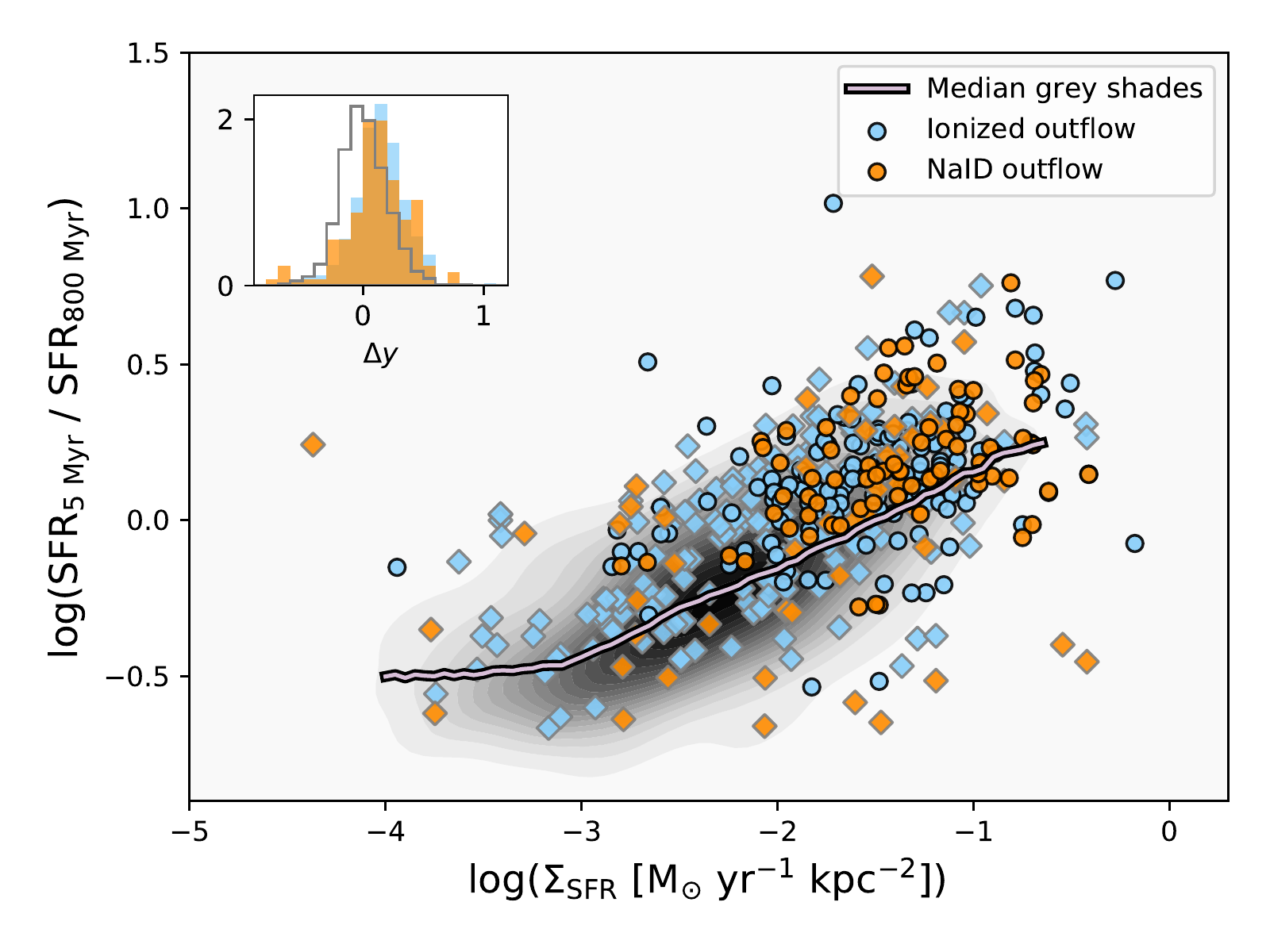}
\vspace{-0.8cm}
\caption{
SFR change parameter versus SFR surface density. The pink line presents the running median of the MaNGA analysed sample whose distribution is shown by grey shades. The positions of galaxies with ionized winds are shown in blue symbols and galaxies with Na I D winds are shown in orange. Circle and diamond symbols denote galaxies without and with nuclear activity, respectively. The histogram in the top-left shows the distribution of the vertical distances between data points and the pink line.}
\label{fig:SFR79}
\end{figure*}

\subsubsection{SFR change parameter}

We now turn to investigate the dependence of outflow detection on recent changes in the star-formation history given that processes associated with star-formation are important drivers of galactic winds in typical nearby galaxies. The motivation for this is that we could imagine a scenario where the ionized winds, which probe the warm wind phase, are more instantaneously related to a recent fluctuation in SF activity upwards, compared to the neutral-gas winds which are colder and therefore may probe winds until later after the initial launching event, imprinted in absorption onto the background light as they are coasting away from their launching site. 

To investigate this, we calculate the SFR change parameter ($\rm SFR_{5 Myr}/SFR_{800 Myr}$) defined by \citet{Wang2020} as the SFR averaged over the past 5 Myr divided by the SFR averaged over the past 800 Myr. This parameter can take values above or below unity depending on whether there was a recent enhancement or suppression of star formation relative to the typical star formation occurring over the past 800 Myr. It is computed using the H$\alpha$ equivalent width, H$\delta$ absorption equivalent width, and strength of the 4000$\Angstrom$ break as inputs, and calibrated using a suite of stellar population synthesis models (see \citealt{Wang2020} for details). These parameters are computed within the same $R_{\rm out}$ sized aperture used to probe the outflow properties.

In Fig. \ref{fig:SFR79} we show the position of objects in the $\rm SFR_{5 Myr}/SFR_{800 Myr}$ versus $\Sigma_{\rm SFR}$ plane, where the former parameter is calculated within the outflow aperture. We find a positive correlation between the two in the underlying analysed sample and the outflow subsamples. The correlation in the analysed sample can be clearly seen in the median trend line (pink line in Fig. \ref{fig:SFR79}), where we use a `sliding window' to calculate the median $\rm SFR_{5 Myr}/SFR_{800 Myr}$ within a given $\Sigma_{\rm SFR}$ interval of width 0.3 dex and in increments of 0.05 dex. The presence of such correlation is not necessarily surprising.  The MPA-JHU star formation rates on which the galaxy-averaged star formation surface densities are based are inferred from H$\alpha$ and thus probe short timescales, making it plausible that at higher $\Sigma_{\rm SFR}$ more galaxies are found to have recently turned up their star formation activity.

The data in the inset histogram in Fig. \ref{fig:SFR79} are collected by measuring the vertical distance of the outflow objects from this median trend. This histogram shows that galaxies with detectable winds in the ionized phase, or in Na I D, preferentially have higher $\rm SFR_{5 Myr}/SFR_{800 Myr}$ values than one might expect given their galaxy star formation surface densities, and we verify this offset is significant using the K-S statistic. Fig. \ref{fig:SFR79} further shows that outflow objects preferentially have $\rm SFR_{5 Myr}/SFR_{800 Myr}$ values above one, particularly those classified as star forming with no active nucleus according to the BPT diagnostic\footnote{We note that the SFR change parameter as introduced by \citet{Wang2020} was calibrated using pure stellar population models, and may therefore be compromised in the presence of AGN contributions to the input spectral diagnostics, especially EW(H$\alpha$).  While galaxies with an active nucleus are for completeness included in Fig.\ \ref{fig:SFR79}, we tested that the same conclusions hold when restricting the samples to inactive galaxies only.}, where $\sim 80$ per cent of star forming systems show $\rm SFR_{5 Myr}/SFR_{800 Myr}$ values $> 1$. This indicates that these objects had a recent upturn in their SFR. 

When comparing the different gas phases, we find no significant offset in the distributions of $\rm SFR_{5 Myr}/SFR_{800 Myr}$ values at a given $\Sigma_{\rm SFR}$. The lack of difference between the ionized and Na I D means we cannot draw conclusions about the relative timescales over which the different tracers probe the outflowing gas. 

We point out to the reader that the 74 objects with outflows detected in both ionized and neutral phases appear twice in figures \ref{fig:dust} and \ref{fig:SFR79} (once as a subsample of the orange points, and once as a subsample of the blue points), since the y-values are based upon measurements made within the outflow aperture (of radius governed by the outflow extent), which is derived separately for the ionized and neutral winds.

\begin{figure}
\centering
\includegraphics[width=0.45\textwidth]{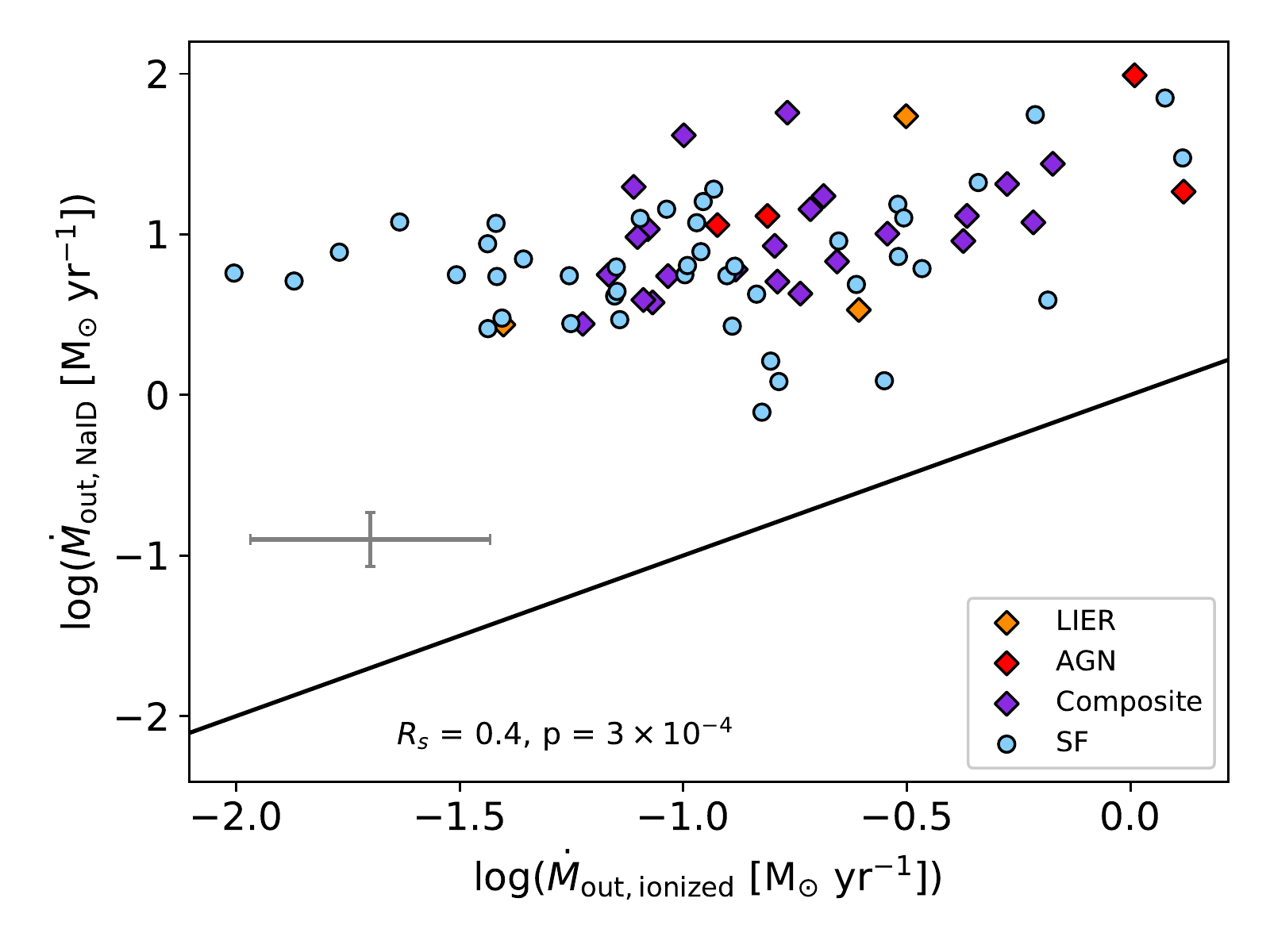}
\includegraphics[width=0.45\textwidth]{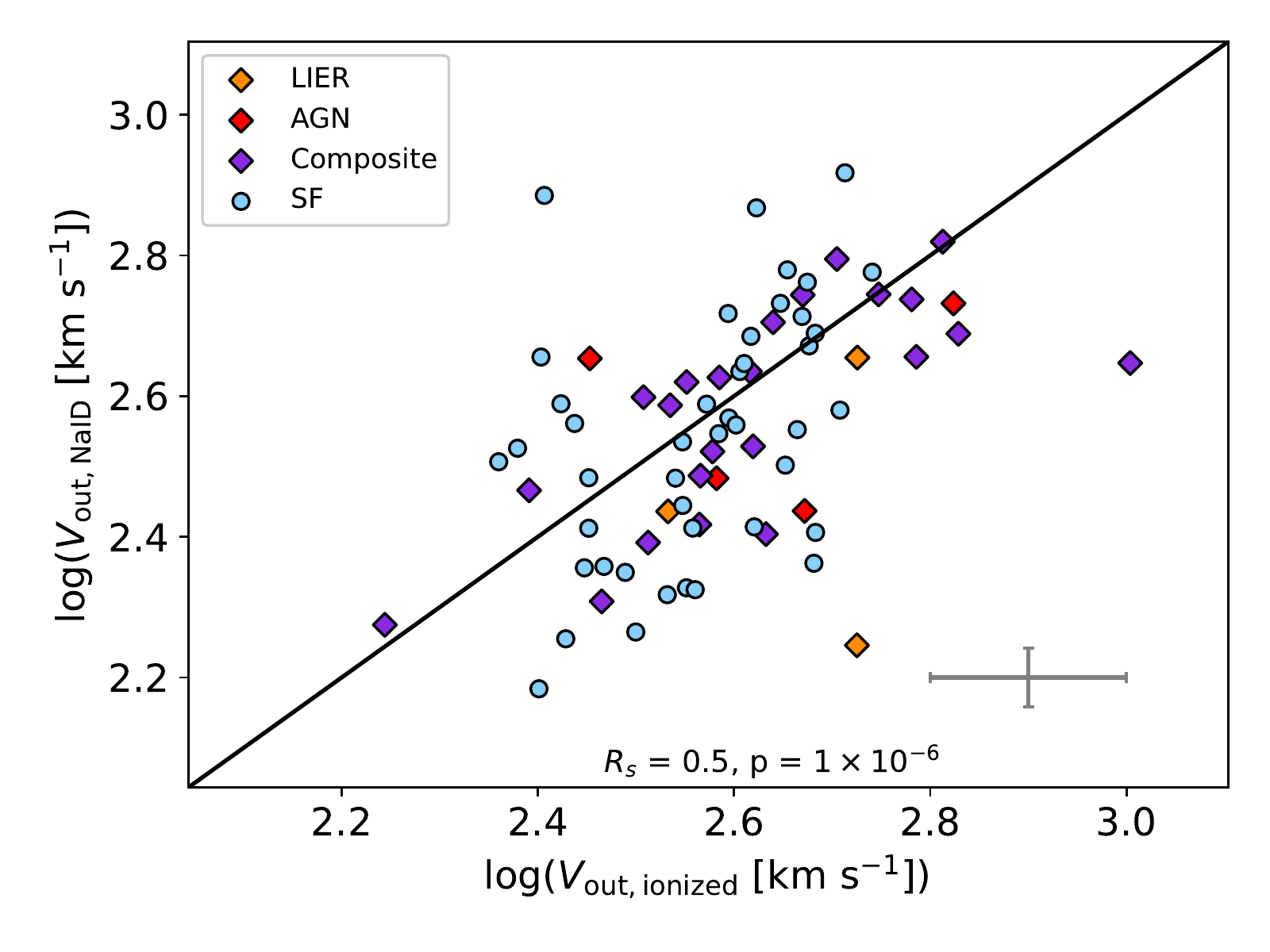}
\caption{Comparison of the mass outflow rate ({\it top}) and outflow velocity ({\it bottom}) in the ionized and neutral gas phases within the same MaNGA objects. The black solid line shows the one-to-one relation. Typical errors are shown.}
\label{fig:phase_comparison}
\end{figure}

\subsection{Ionized versus neutral gas outflows}
\label{phases.sec}

In this section we directly compare the outflow properties measured in the cool neutral and warm ionized phases.  In Fig. \ref{fig:phase_comparison}, we present a comparison for $\dot{M}_{\rm out}$ and $v_{\rm out}$, for the subset of 74 MaNGA galaxies where winds are detected in both Na I D and ionized emission.

We find $\dot{M}_{\rm out}$ is correlated between the two phases with a Spearman's rank correlation coefficient $R_s \sim 0.4$ ($p \sim 3 \times 10^{-4}$). Mass outflow rates are significantly larger in the neutral gas phase compared to the ionized phase, with a median offset of $\sim 1.8$ dex. This latter result is roughly consistent with other empirical studies on multiphase outflows including \citet{RobertsBorsani2020} and \citet{Fluetsch2021}.

In \citet{Avery2021}, we identified clues that the outflowing wind medium may be dust-enhanced relative to the ISM in the galaxy disks, expressed by an elevated Balmer decrement for the broad-component line emission.  If taking these (challenging) measurements at face value, and applying dust correction factors based on the broad-component H$\alpha$/H$\beta$, rather than the default Balmer decrements from single-Gaussian fits (predominantly probing the light from the galaxy disk), the ionized mass outflow rates go up, but remain systematically below those in the neutral gas phase, with a median offset of $\sim 1.2$ dex.  

One may further note in Fig.\ \ref{fig:phase_comparison} that the $\dot{M}_{\rm out,NaID}/\dot{M}_{\rm out,ionized}$ ratio appears to increase among weaker winds.  We caution against overinterpreting this trend, in the sense that not for all galaxies with ionized winds a Na I D outflow signature was detectable.  Among weak (and less obscured) ionized winds, a counterpart $\dot{M}_{\rm out,NaID}$ measurement is more frequently missing, and it may therefore be that those objects featuring on the left side of Fig.\ \ref{fig:phase_comparison} ({\it top panel}) represent the upper tail of a distribution in $\dot{M}_{\rm out,NaID}$.

Broadening our perspective to all wind phases, \citet{Fluetsch2021} have shown that in local ULIRGs, the ionized wind component often only makes up a very small (or even negligible) fraction of the total outflow mass budget, whilst the neutral phase contributions are of the order 10 per cent, and the molecular phase can account for up to $\sim 95$ per cent of the outflow mass in some objects presented in their study. \citet{RobertsBorsani2020} find similar results in the more typical galaxy population where they show via a stacking procedure that the molecular phase accounts for the majority of the outflowing material, whilst the neutral phase accounts for around 30 per cent, and the ionized phase is negligible. 

The $v_{\rm out}$ comparison shown in the bottom panel of Fig. \ref{fig:phase_comparison} shows a correlation between the outflow velocities in the ionized and neutral phases along the one-to-one line ($R_s \sim 0.5, p \sim 1\times10^{-6}$). This indicates that despite the fact that the ionized gas only makes up a fraction of the wind, its kinematics are characteristic of the motion of the neutral gas which makes up a larger fraction of the outflowing material. Separating objects by their excitation type, as assessed via the [NII]-BPT diagram, we find this trend to hold for both non-active and active galaxies.  Outflow velocities span the range $v_{\rm out} \sim 190 - 640 \rm \ km \ s^{-1}$ (median $v_{\rm out} = 363 \rm \ km \ s^{-1}$) for Na I D winds, and for the ionized phase $v_{\rm out} \sim 250 - 620 \rm \ km \ s^{-1}$ (median $v_{\rm out} = 389 \rm \ km \ s^{-1}$). Comparing to the literature, \citet{Concas2019} find their most extreme Na I D wind velocities to be within the upper half of the range seen here. Furthermore, previous studies tend to find typically slower wind velocities in the neutral phase compared to the ionized phase \citep[e.g.,][]{RobertsBorsani2020, Baron2021}.  In this context, we note that if we were to define the neutral outflow velocity by the 95th bluest percentile of the gaseous Na I D absorption (as e.g. adopted in \citealt{Baron2021}) rather than by Eq.\ \ref{eq:vout}, the amount by which ionized winds, in the median, are faster increases from 0.03 dex to 0.13 dex.

We further find the ionized outflow to be more extended than the Na I D outflow with 55 (out of 74) objects having smaller calculated outflow radii in Na I D. The median of the distribution of $R_{\rm out}$ values for the ionized and neutral outflow samples is 3.8 and 1.9 kpc, respectively. Central concentrations of dust as revealed by the typically negative gradients of dust attenuation within galaxies may be at least partially responsible for the centrally concentrated nature of Na I D outflows.

\subsection{Scaling relations for Na I D outflows}
\label{scaling.sec}

\begin{figure}
\centering
\includegraphics[width=0.45\textwidth]{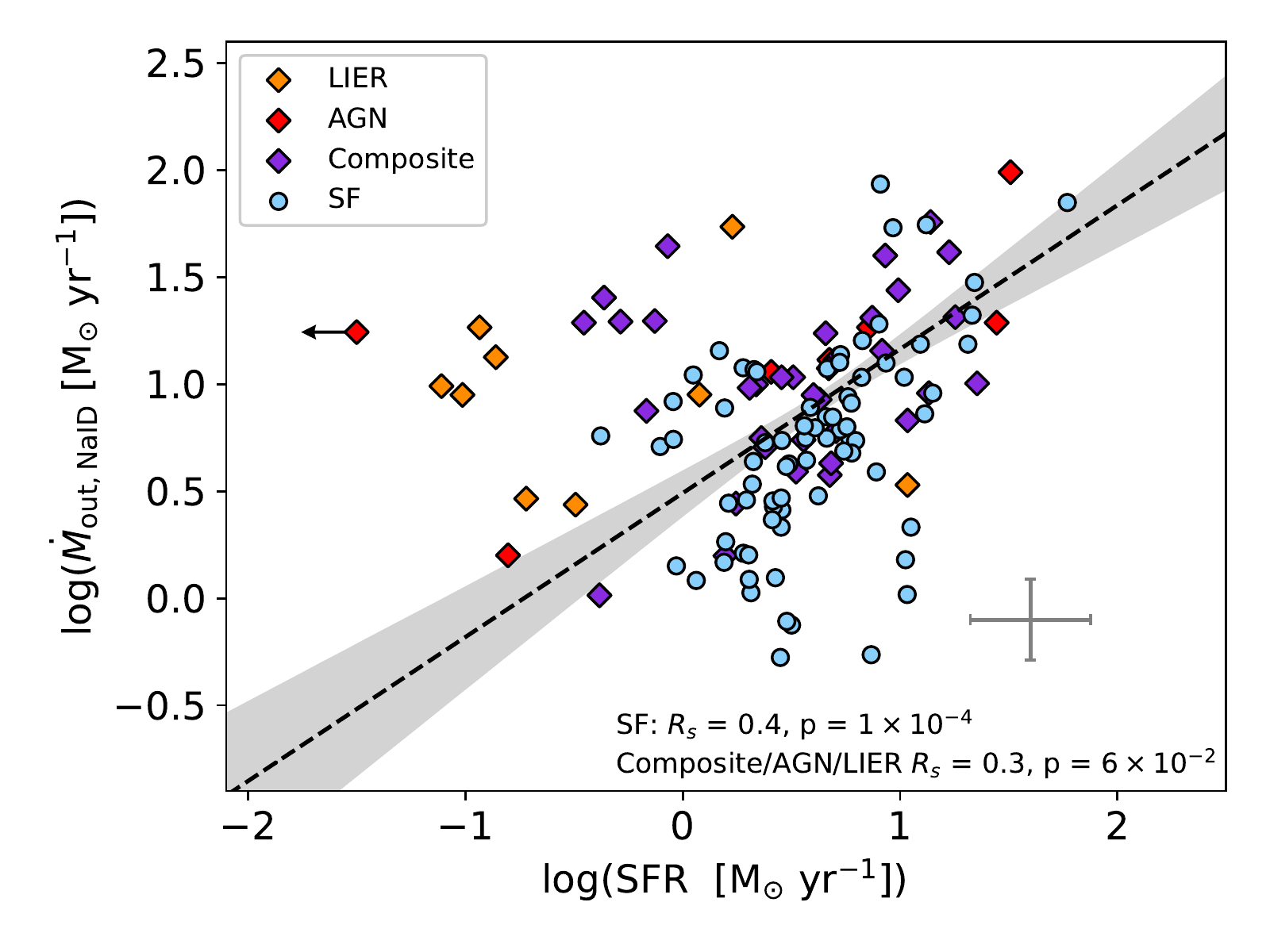}
\includegraphics[width=0.45\textwidth]{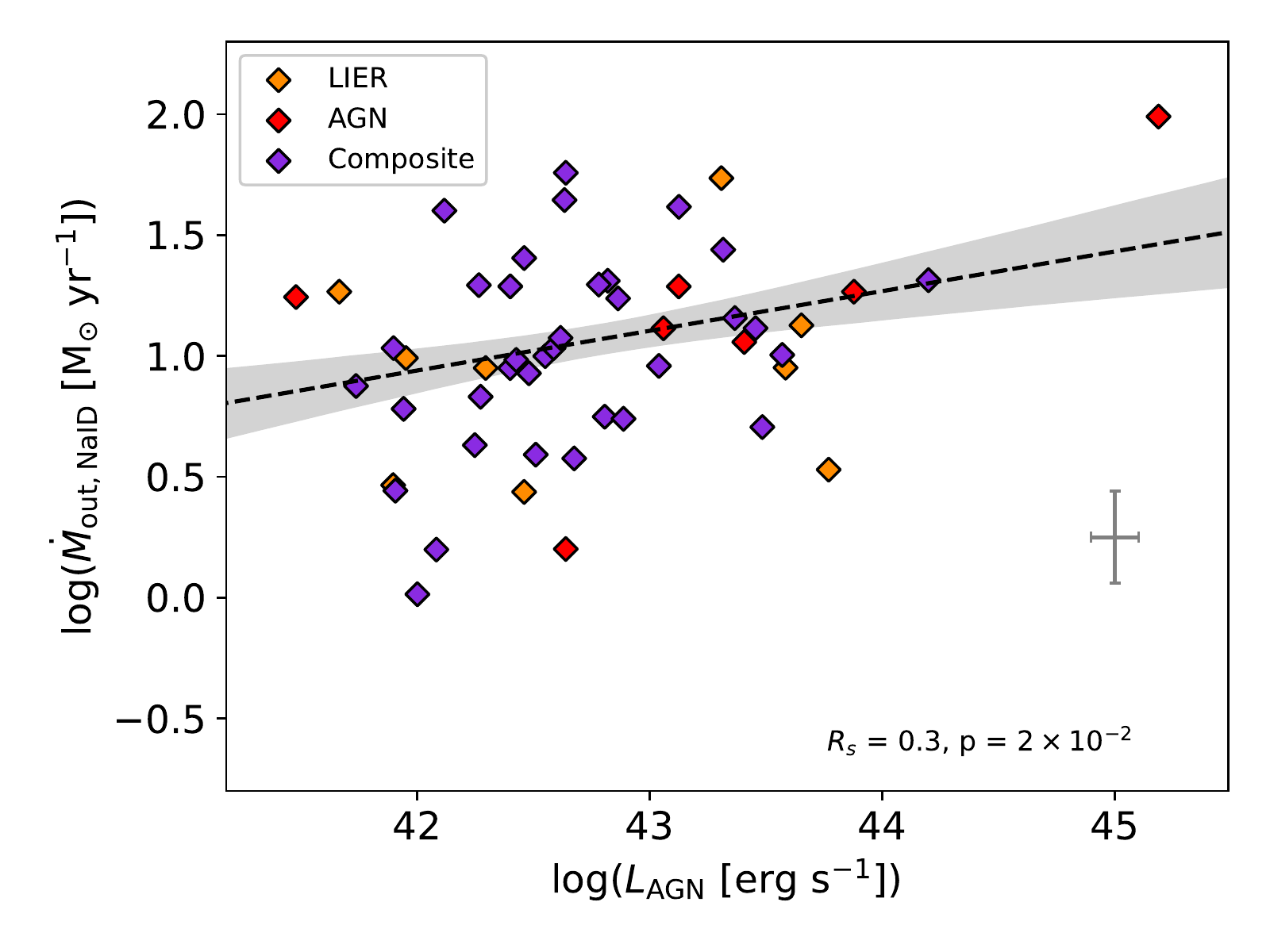}
\caption{{\it Top:} Neutral gas mass outflow rate as a function of SFR. {\it Bottom:} Neutral gas mass outflow rate as a function of AGN luminosity for active systems. The grey shaded region indicates the 1$\sigma$ error on the best fitting linear trend (black dashed line). Typical errors are shown. Spearman's rank correlation coefficients (and their associated p-values) are listed for inactive and active galaxies separately.}
\label{fig:Mdot_out_drivers}
\end{figure}

In this section we present and discuss the scaling relations uncovered between the strength of neutral gas outflows, characterised by $\dot{M}_{\rm out}$, and the strength of their physical drivers SFR and $L_{\rm AGN}$. These relations are shown in Fig \ref{fig:Mdot_out_drivers}.

The python package \texttt{linmix} \citep{Kelly2007} is used to determine linear fits to the relations as it takes into account the errors in both x and y. The linear regression in $\dot{M}_{\rm out} -$ SFR space gives a power law slope of $0.67 \pm 0.1$ with data showing a scatter of 0.56 dex about the trend line. Active galaxies tend to sit above the bestfit line, plausibly due to contributions from nuclear activity besides star formation to the wind driving, and account for more outliers in the relation. This is reflected in the smaller scatter ($\sim 0.42$ dex) and steeper slope ($0.74 \pm 0.2$) measured when they are removed from the top panel of Fig. \ref{eq:Mdot_out}. When considering non-active systems alone, we find power law slopes in agreement (at the 1$\sigma$ level) for the ionized and Na I D phases, where we found $\dot{M}_{\rm out} \propto \rm SFR^{0.97 \pm 0.07}$ for the ionized winds. When considering all systems, the $\dot{M}_{\rm out} -$ SFR relation derived using the Na I D feature has a $\sim 0.1$ dex larger scatter than the ionized relation presented in \citet{Avery2021}. 

Comparing to the literature, \citet{RobertsBorsani2019} performed a stacking analysis on SDSS single-fibre spectra and found a steeper relation of the form $\dot{M}_{\rm out, NaID} \propto \rm SFR^{1.08 \pm 0.17}$, albeit with a lower zero-point, when considering both non-active and active galaxies. Although their slope measurement is only in agreement with our result at the 2$\sigma$ level, our study extends outflow measurements down to lower SFRs ($< 1 \rm M_{\odot} yr^{-1}$) where we see active galaxies sitting above the trend line. In limiting the SFR range to that explored by \citet{RobertsBorsani2019}, we find a slope in agreement at the $1\sigma$ level. When considering the absolute $\dot{M}_{\rm out}$ values, we find a median $\dot{M}_{\rm out} \sim 7.3 \ \rm M_{\odot}yr^{-1}$ which is consistent with other neutral outflow studies in typical low-redshift galaxies \citep{RobertsBorsani2020}, and the post-starburst galaxies studied by \citet{Baron2021}.

Turning to the subset of active galaxies, we find a weak trend between $\dot{M}_{\rm out}$ and $L_{\rm AGN}$, where the power law slope defining the relation is given by $0.16 \pm 0.1$, and scatter 0.39 dex. This relation is significantly shallower than found for the ionized gas in \citet{Avery2021}, with a similar scatter.  We note that, relative to the active galaxies in the ionized outflow sample, the active galaxies with Na I D winds comprise a modestly larger proportion of so-called "Composite" systems (lying between the \citealt{Kauffmann2003} and \citealt{Kewley2001} curves on the [NII]-BPT diagram). Here we use AGN luminosities derived from [OIII] emission which are less subject to short-timescale variability of the AGN than X-rays.  Nevertheless, nuclear variability over timescales shorter than those over which winds signatures are seen can contribute to scatter in this diagram. 

\begin{figure*}
\centering
\includegraphics[width=\textwidth]{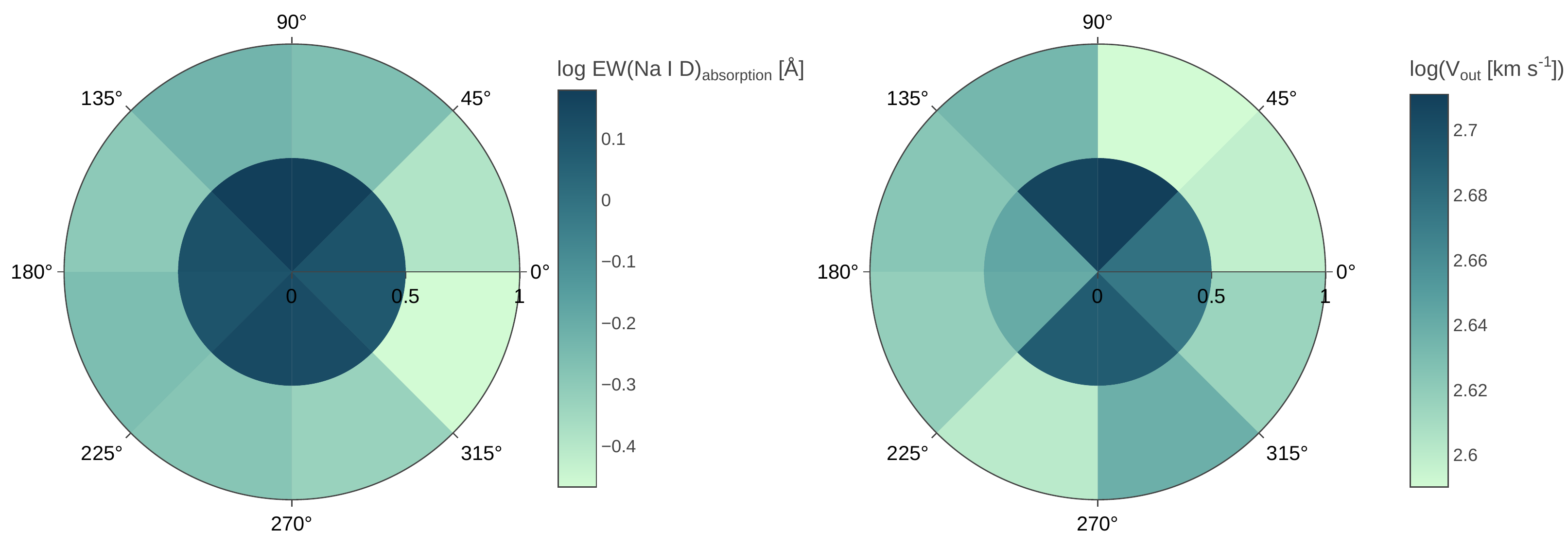}
\caption{Average outflow properties derived from stacking in spatial bins across the neutral outflow sample. Stacking results are displayed over the radial range 0 - 1 $R_{\rm e}$, and with azimuthal binning defined such that the orientation of galaxy major axes runs horizontally on the diagrams. The radial axis has units of $R_{\rm e}$. Outflows were undetected at a radius larger than $1 R_{\rm e}$ in the stacks. {\it Left:} Average EW of the gaseous NaID absorption feature across the galaxies. {\it Right:} Average neutral gas outflow velocity. }
\label{fig:polar_plots}
\end{figure*}

As well as $\dot{M}_{\rm out}$, we also consider the mass-loading ($\eta$) in the neutral phase, defined as $\dot{M}_{\rm out}$ divided by the total galaxy SFR. This is an important parameter in galaxy evolution models informing on the efficiency of star formation driven winds. We find, with all caveats of, e.g., assumed covering factors etc., that the obtained wind mass loading factors are of order unity for the inactive galaxies, and within the $10 \lesssim \rm log(M_{\star}) \lesssim 11$ dynamic range do not show an obvious dependence on stellar mass. The lack of strong negative dependence of mass loading on $M_{\rm \star}$, as invoked indirectly to account for the shape of the stellar mass - metallicity relation \citep[see, e.g.,][]{Lilly2013}, may be attributed to the fact that our MaNGA outflow sample does not extend far enough down in mass, or alternatively may imply that there is an important distinction between the amount of outflowing material being observed via direct tracers compared to the effective mass loading in gas regulator models which encodes how much of the gas reservoir is entirely lost from the galaxy system. Indeed, we find that the ratio of outflow velocity to escape velocity $v_{\rm out}/v_{\rm esc}$ (where $v_{\rm esc}$ is estimated as outlined in Section\ 4.4 of \citealp{Avery2021}) increases towards lower masses, suggesting that outflows may be more likely to escape from the lower mass systems within our sample.  Overall, the fraction of objects for which this ratio exceeds unity remains limited to $\sim 25$ per cent of the neutral gas outflows.  Inferred mass loading factors among galaxies hosting an active nucleus extend to higher, above-unity values, as may be anticipated from the fact that some of them are found in galaxies residing below the star-forming main sequence.

\subsection{Outflow geometry}
\label{geometry.sec}

\begin{figure}
\centering
\includegraphics[width=0.95\linewidth]{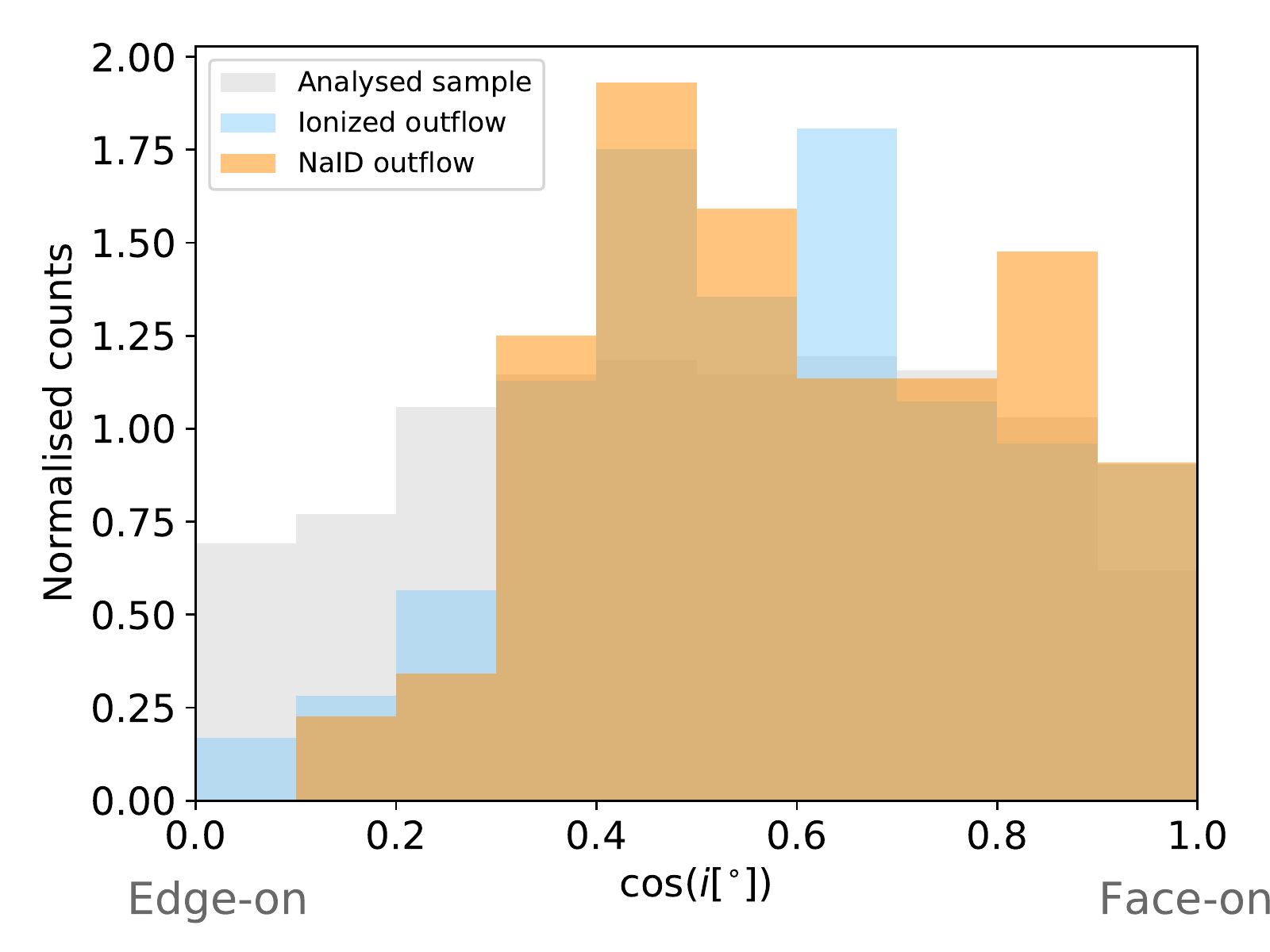}
\includegraphics[width=0.98\linewidth]{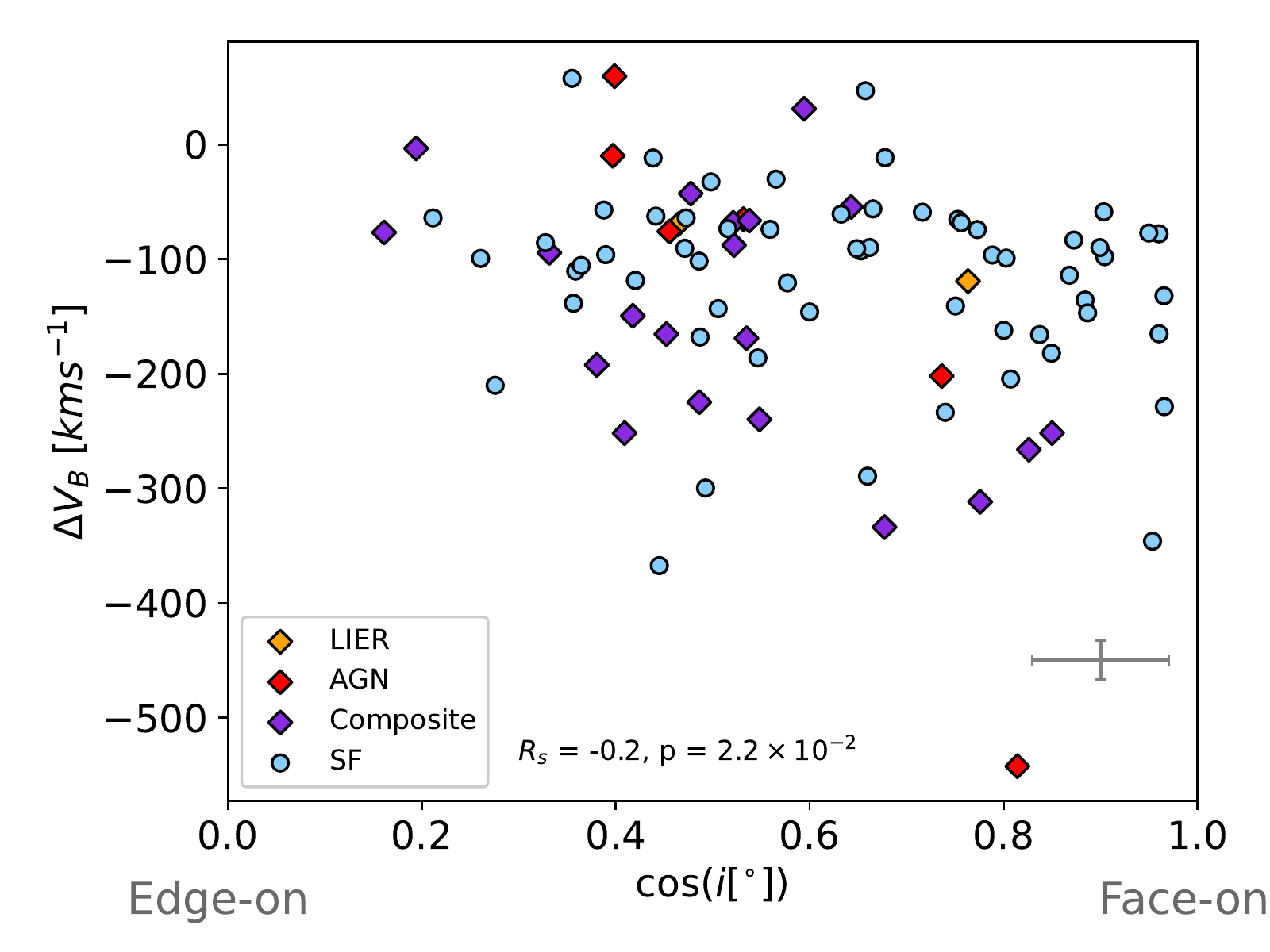}
\caption{{\it Top:} Distribution of galaxy inclinations for the underlying analysed sample (grey), the ionized wind sample (blue) and the Na I D wind sample (orange). We find that winds are preferentially detected in systems away from edge-on orientations.  {\it Bottom:} Velocity offset of the outflow component in Na I D relative to the galaxy's systemic velocity as a function of cosine inclination.}
\label{fig:incl}
\end{figure}

To achieve an insight into the structure of the cool gas outflows, we performed a radial-azimuthal binning + stacking technique on our neutral outflow sample. For each object in the neutral outflow sample, we extract the \texttt{SPX\_ELLCOO} extension of the associated MaNGA MAPS file, which provides the elliptical polar coordinates of each spaxel from the galaxy center. Using this, we bin the spaxels in the MaNGA data cube into four radial bins in units of $R_{\rm e}$: $0-0.5$; $0.5-1$; $1-1.5$; $>1.5$, and then subdivide each radial bin into eight azimuthal bins, each of width $45^{\circ}$. Azimuthal bins are created such that the first azimuthal bin for each galaxy covers spaxels 0 to $45^{\circ}$ away from the galaxy's major axis. We next summed the spectra of spaxels within a given bin after removing the gaseous velocity field and interpolating spectra onto a common velocity grid. We applied a median normalization to the binned spectra, and subsequently combined these normalized spectra belonging to a particular radial-azimuthal bin from the different objects in the outflow sample. We then repeated the methodology outlined in sections \ref{sec:NaID_fitting} and \ref{sec:criteria} to perform a decomposition on the Na I D feature in each bin and determine whether or not an outflow component is detected. Using the best-fitting parameters, we estimate the outflow velocity and integrate over the gaseous ${\rm Na\ I\ D}$ absorption component, originating from the ISM and outflow, to measure the equivalent width (EW). The results are presented in Fig. \ref{fig:polar_plots}. We find the outflows to be centrally concentrated, where outflows beyond 1 $R_{\rm e}$ are too weak and/or only appear in a minority of objects, such that they do not appear in our radial-azimuth stacks. Looking at the EW, on average, we see stronger absorption in the central region with the EW being highest within 0.5$R_{\rm e}$ and decreasing out to 1 $R_{\rm e}$.

Turning to the $v_{\rm out}$ map, a similar trend of central concentration is notable, and we see that the average $v_{\rm out}$ for the wind sample is highest at angles perpendicular to the major axis of the galaxy (within 0.5$R_{\rm e}$). This is consistent with the readily assumed scenario of biconical outflows being driven along the minor axis of the galaxy, orthogonal to the disk plane. 

Other studies of outflow detection in absorption have highlighted that outflows can be more readily detected in systems with orientations away from edge-on \citep[e.g.,][]{Heckman2000}. 

In Fig. \ref{fig:incl}, we investigate inclination dependencies for a subsample of the neutral outflow galaxies which are star-forming (log(sSFR) $> -11$) and have disc morphologies\footnote{Disc morphologies are defined as galaxies which are not classified as ‘odd looking’, or appear to be undergoing a merger according to the Galaxy Zoo 2 \citep{GalaxyZoo2} and the Galaxy Zoo 1 \citep{Lintott2011} classifications, respectively.}, such that an estimate of inclination can reliably be derived from the galaxy axial ratio as follows:
\begin{equation}
    \mathrm{cos}(i) = \sqrt{\dfrac{(b/a)^2 - \mathrm{thickness}^2}{1 - \mathrm{thickness}^2}}
\label{eq:incl}
\end{equation}
Here, $b/a$ is the galaxy’s semi-minor-to-semi-major axial
ratio, and is given as output from the MaNGA data analysis pipeline. A characteristic disc thickness (i.e. ratio of scale height over scale length) of 0.15 is assumed, appropriate for thin nearby disc galaxies. We estimate the error on cos($i$) by propagating the error on $b/a$, thus assuming the latter to serve as a reliable proxy for inclination. We show, from Fig. \ref{fig:incl} that the distribution of inclinations for galaxies exhibiting neutral outflows is consistent with the result for outflows in the ionized phase, where both wind phases are indeed preferentially detected in systems with orientations away from edge-on. Furthermore, we find a significant, albeit weak ($R_s \sim 0.2$, $p \sim 2.2\times 10^{-2}$), correlation between inclination and the amplitude of the velocity offset of the outflow component relative to the systematic gas in the galaxy disk. This trend is also seen in other studies of outflows probed in absorption \citep[e.g,][]{Concas2019} and hints towards the biconical outflow model. We evaluate whether this correlation holds when measurement errors are taken in account by perturbing the cos($i$) and $\Delta v_{\rm B}$ values randomly within a normal distribution, centered on the measured value, and of width equal to the associated error. We do this 1000 times, and each time evaluate the $R_s$ and $p$ values. We find that $\sim 50$ per cent of the bootstrap realisations recover a statistically significant correlation ($p < 0.05$)\footnote{Technically, the quoted statistic from the bootstrap test represents a worst case scenario, as an additional perturbation by the measurement uncertainties is applied, on top of the measurement values which themselves already scatter around the true, intrinsic quantities.}.

\section{Summary and Conclusions}
\label{summary.sec}

In this paper, we have characterised the incidence and strength of cool neutral winds among line-emitting galaxies in the MaNGA galaxy survey. This work complements our previous investigation of the ionized gas winds in the same sample of galaxies \citep{Avery2021}. We have directly compared the strength of the outflows in two gas phases, building upon the literature by extending the sample to the more `normal' galaxy population in the nearby Universe whilst not losing information about individual objects due to stacking.

Our main results are:
\begin{itemize}
    \item Na I D wind detection appears to be dependent on the amount of dust in the host galaxy environment (Fig. \ref{fig:dust}). This effect can largely explain why we observe neutral gas winds in galaxies with higher $\Sigma_{\rm SFR}$, compared to the incidence of ionized gas winds (due to the correlation between $\Sigma_{\rm SFR}$ and dust attenuation).  Additionally, the brighter background continuum source in high-$\Sigma_{\rm SFR}$ galaxies may contribute to making the Na I D outflow absorption signature more visible. 
    \item We present evidence that wind galaxies frequently have had a recent upturn in their SFRs compared to the underlying MaNGA sample (Fig. \ref{fig:SFR79}) according to the SFR change parameter introduced by \citet{Wang2020}.
    \item There is a positive correlation between the mass outflow rates measured in the ionized and neutral gas phases, with the neutral winds having higher typical mass outflow rates by $\sim 1.2 - 1.8$ dex, depending on the dust correction applied (Fig. \ref{fig:phase_comparison}).
    \item An approximate one-to-one positive correlation between the outflow velocities in the two phases indicates that the ionized gas traces the neutral gas kinematics (Fig. \ref{fig:phase_comparison}). 
    \item A positive correlation between $\dot{M}_{\rm out}$ and SFR is observed for Na I D wind galaxies, and to a lesser extent between $\dot{M}_{\rm out}$ and $L_{\rm AGN}$ for the subset featuring nuclear activity (Fig. \ref{fig:Mdot_out_drivers}). 
    \item By sectioning the wind galaxies into radial and azimuthal segments, and stacking segments across the neutral outflow sample, we create spatial maps of the depth of the Na I D absorption and the outflow velocity (Fig. \ref{fig:polar_plots}). We find that outflows are, on average, centrally concentrated, and can escape their launching sites fastest along the galaxy minor axis. From this, and further corroborated by inclination dependencies of outflow incidence and the blueshift of Na I D absorption (Fig. \ref{fig:incl}), we find consistency with a picture of winds which are biconical and perpendicular to the disk plane.
\end{itemize}

Although ionized gas studies may not capture the bulk of outflowing material, we show that they are valuable because the outflow kinematics are in line with what we find in the neutral phase which carries a more significant fraction of the overall outflow mass budget. Furthermore, we find that studies employing ionized gas tracers are more complete in picking up signatures of feedback in action across a wider range of galaxy properties, unlike Na I D detected winds which appear to have a requirement for dusty environments. 

Further progress in this area would benefit from a systematic, matched-resolution follow-up of sizeable numbers of MaNGA galaxies in molecular gas tracers.  Such millimetre-wavelength datacubes of low/mid-J CO transitions could complement the ionized and neutral gas wind diagnostics by quantifying the molecular contribution to the wind medium, but would equally yield insight on the winds' (localized) impact on the galaxies' cold gas reservoirs.

\section*{Acknowledgements}
We thank the referee for their helpful and insightful comments which
have motivated several improvements to this paper.

Funding for the Sloan Digital Sky Survey IV has been provided by the Alfred P. Sloan Foundation, the U.S. Department of Energy Office of Science, and the Participating Institutions. SDSS-IV acknowledges support and resources from the Center for High-Performance Computing at the University of Utah. The SDSS web site is www.sdss.org.

SDSS-IV is managed by the Astrophysical Research Consortium for the  Participating Institutions of the SDSS Collaboration including the 
Brazilian Participation Group, the Carnegie Institution for Science,  Carnegie Mellon University, the Chilean Participation Group, the French Participation Group, Harvard-Smithsonian Center for Astrophysics, Instituto de Astrof\'isica de Canarias, The Johns Hopkins University, Kavli Institute for the Physics and Mathematics of the Universe (IPMU) / University of Tokyo, the Korean Participation Group, Lawrence Berkeley National Laboratory, Leibniz Institut f\"ur Astrophysik Potsdam (AIP), Max-Planck-Institut f\"ur Astronomie (MPIA Heidelberg), Max-Planck-Institut f\"ur Astrophysik (MPA Garching), Max-Planck-Institut f\"ur Extraterrestrische Physik (MPE), National Astronomical Observatories of China, New Mexico State University, New York University, University of Notre Dame, Observat\'ario Nacional / MCTI, The Ohio State University, Pennsylvania State University, Shanghai Astronomical Observatory, United Kingdom Participation Group,
Universidad Nacional Aut\'onoma de M\'exico, University of Arizona,  University of Colorado Boulder, University of Oxford, University of Portsmouth, University of Utah, University of Virginia, University of Washington, University of Wisconsin, Vanderbilt University, and Yale University.

\section*{Data Availability}
The data underlying this article were accessed from the publicly available Data Release 15 (DR15) of the Sloan Digital Sky Survey (SDSS-IV). Specifically we make use of the integral field unit (IFU) spectroscopic observations of nearby galaxies from the SDSS Mapping Nearby Galaxies at APO (MaNGA) survey.  All of the MaNGA data can be accessed using the traditional SDSS Science Archive Server (SAS) at https://data.sdss.org/sas/dr15/manga/spectro/. The NASA-Sloan Atlas can be accessed at http://nsatlas.org/.
The processed datasets underlying this article, which were derived as outlined in Section \ref{method.sec}, will be shared on reasonable request to the corresponding author.



\bibliographystyle{mnras}
\bibliography{ms} 




\appendix
\onecolumn

\section{Properties of host galaxies and their outflows.}
\label{sec:appendix}

\begin{figure}
\centering
\includegraphics[width=\textwidth]{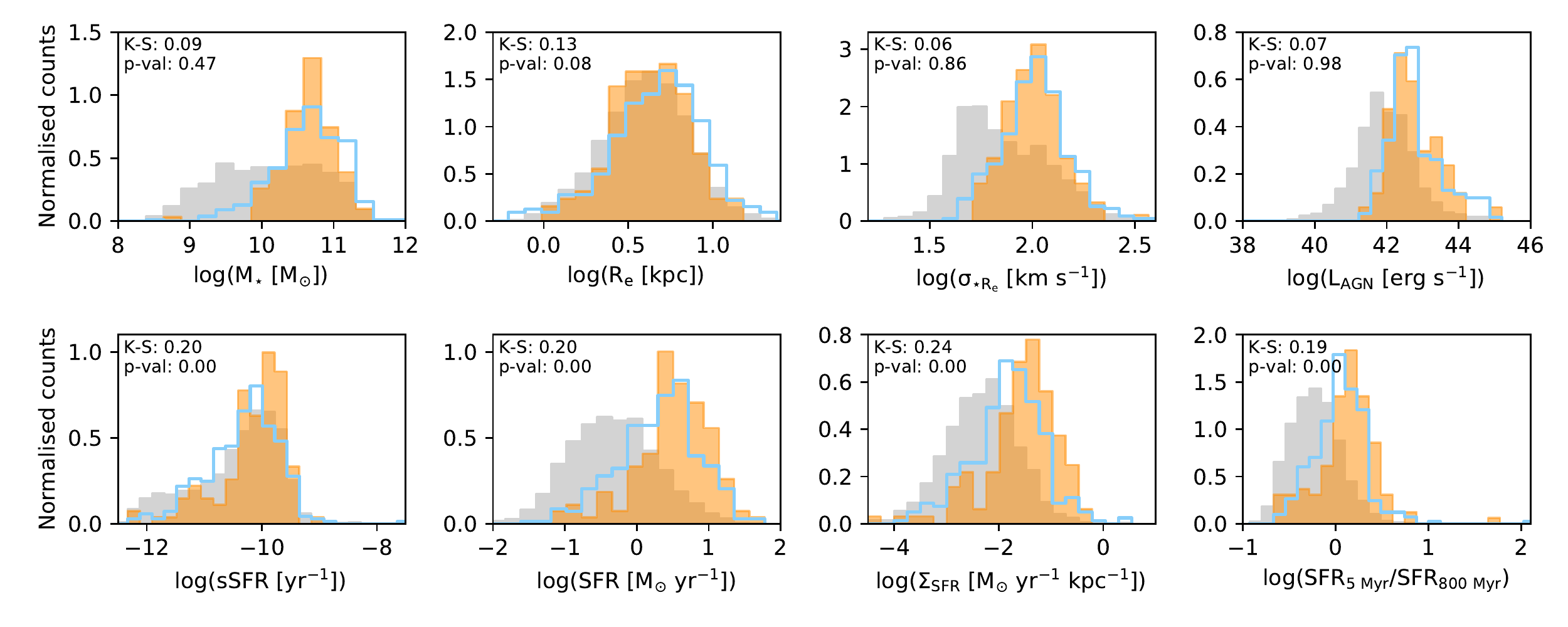}
\caption{Distribution of the underlying analysed (grey), ionized wind (blue), and Na I D wind (orange) samples among various galaxy properties. Kolmogorov-Smirnov test values and their associated p-values comparing ionized wind galaxies and neutral wind galaxies are shown. AGN luminosities ($L_{\rm AGN}$) are shown only for those galaxies with identified AGNs.}
\label{fig:incidence}
\end{figure}

In Fig. \ref{fig:incidence}, we compare the distributions of galaxy properties among the ionized outflow sample \citep{Avery2021}, the neutral outflow sample (this work), and the underlying analysed MaNGA sample of line-emitting galaxies. We further provide a sample table listing host galaxy and outflow properties for the neutral and ionized outflow samples (Table \ref{tab:all_MaNGA_outflows}; full table provided online).

\begin{table} 
\centering 
\resizebox{\linewidth}{!}{%
\begin{tabular}{l c c c c c c c c c c c c} 
\hline 
MaNGA ID & $z$ & log($R_{\rm e}$) & $i$ & log($M_{\star}$) & log(SFR) & log($L_{\rm AGN}$) & $\Delta v_{\rm B, ion}$ & $\Delta v_{\rm B, NaID}$ & $v_{\rm out, ion}$ & $v_{\rm out, NaID}$ & log($\dot{M}_{\rm out, ion}$) & log($\dot{M}_{\rm out, NaID}$)\\
 &  & [kpc] & [$^{\circ}$] & $[\rm M_{\odot}]$ & $[\rm M_{\odot} \ yr^{-1}]$ & $[\rm erg \ s^{-1}]$ & $[\rm km \ s^{-1}]$ & $[\rm km \ s^{-1}]$ & $[\rm km \ s^{-1}]$ & $[\rm km \ s^{-1}]$ & $[\rm M_{\odot}\ yr^{-1}]$ & $[\rm M_{\odot}\ yr^{-1}]$\\
\hline 
7443-9101 & 0.041 & 0.67 & 40 & 10.25 & 0.47 & 42.34 & -65.8 & - & 297.9 & - & -1.62 & - \\
7815-3702 & 0.030 & 0.36 & 32 & 10.37 & 0.43 & - & - & -182.0 & - & 302.6 & - & 0.10 \\
7815-3704 & 0.072 & 0.61 & - & 10.88 & -0.41 & 41.62 & -45.6 & - & 426.3 & - & -2.00 & - \\
7958-6101 & 0.024 & 0.46 & 49 & 10.43 & 0.33 & - & -83.7 & -289.3 & 550.5 & 597.2 & -1.42 & 1.07 \\
7962-9101 & 0.106 & 0.95 & 26 & 11.25 & 1.06 & - & -35.6 & - & 506.2 & - & -0.83 & - \\
7968-9101 & 0.050 & 0.85 & - & 11.17 & -0.07 & 42.63 & - & 19.5 & - & 390.1 & - & 1.64 \\
7972-6103 & 0.044 & 0.68 & - & 11.09 & -1.01 & 42.86 & -38.1 & - & 654.1 & - & -0.71 & - \\
7990-1902 & 0.031 & 0.32 & 56 & 10.73 & 0.72 & - & -7.6 & -30.2 & 308.4 & 223.6 & -0.51 & 1.10 \\
7990-6104 & 0.027 & 0.53 & 32 & 10.03 & 0.38 & - & -61.1 & - & 389.8 & - & -1.46 & - \\
7991-3702 & 0.027 & 0.38 & 52 & 10.42 & -0.38 & 42.60 & -27.6 & - & 326.1 & - & -2.26 & - \\
\hline  
\end{tabular}  
}  
\caption{Sample table displaying data on the properties of MaNGA galaxies with detectable ionized and neutral winds. Spectroscopic redshifts ($z$), and effective radii ($R_{\rm e}$) quantified on the NASA-Sloan Atlas (NSA) $r$-band images, are taken directly from MaNGA DR15. Inclinations ($i$) are derived from the NSA projected axial ratio measurements using Equation \ \ref{eq:incl}, unless morphological classifications (e.g., mergers) suggest the axial ratio may represent a poor proxy of inclination (see Section \ref{geometry.sec}). Galaxy stellar masses and SFRs are adopted from the MPA-JHU database. AGN luminosities are quantified on the basis of the [OIII] luminosity of spaxels with line ratios outside the star-forming region of the BPT-NII diagram (see \citealp{Avery2021} Section 2.2.5 for details). Outflow properties include any systematic blue-shift ($\Delta v_{\rm B}$), the overall outflow velocity ($v_{\rm out}$) and the mass outflow rate based thereupon ($\dot{M}_{\rm out}$). Ionized wind properties (denoted with the subscript `ion') are derived in \citet{Avery2021}, and neutral wind properties (denoted with the subscript `NaID') in this work. The full table can be found online.}
\label{tab:all_MaNGA_outflows}
\end{table}

\section{Relationship between Na I D equivalent width and dust content}
\label{sec:appendixB}

\begin{figure}
\centering
\includegraphics[width=0.5\textwidth]{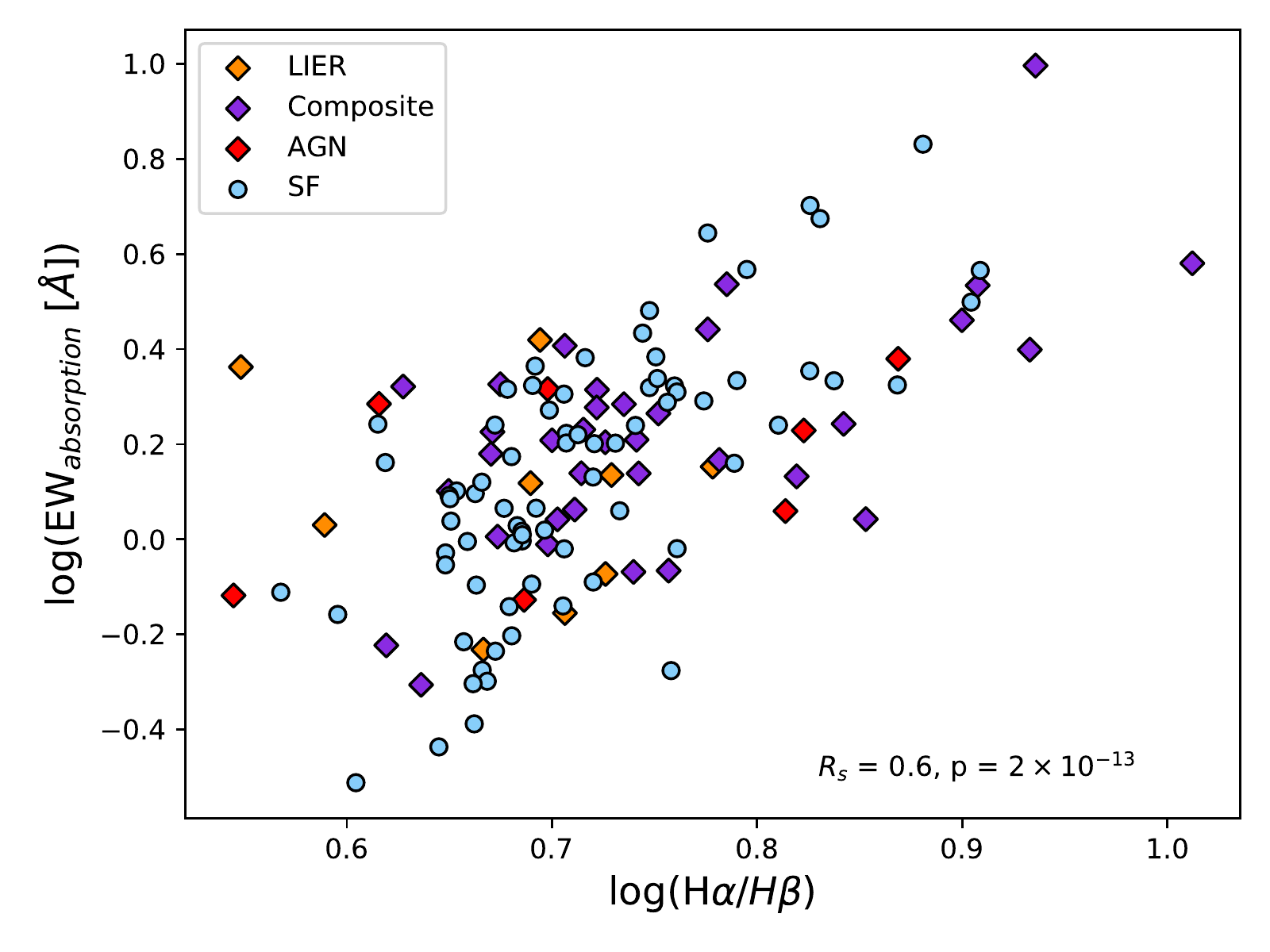}
\caption{Relationship between the equivalent width of the Na I D absorption component and the Balmer decrement within the outflow aperture, for MaNGA galaxies with neutral gas outflows.}
\label{fig:EW_dust}
\end{figure}

The incidence of neutral gas outflows increases among galaxies featuring higher levels of dust attenuation as encoded by the Balmer decrement (Figure\ \ref{fig:dust}).  Not only that, also among the sample with detected neutral gas winds a significant correlation between the equivalent width of the Na I D absorption component and the Balmer decrement is clearly present (Figure\ \ref{fig:EW_dust}).

\bsp	
\label{lastpage}
\end{document}